\documentclass[twocolumn,trackchanges]{aastex7}
\usepackage{multirow}
\usepackage{gensymb}
\usepackage{blindtext}
\usepackage[bottom]{footmisc}
\usepackage{hyperref}
\usepackage{xcolor}
\usepackage{float}  
\usepackage{array}
\newcolumntype{C}[1]{>{\centering\arraybackslash}p{#1}} 
\usepackage{xurl} 

\newcommand{\asec}[1]{{#1}$^{\prime\prime}$}
\newcommand{\amin}[1]{{#1}$^{\prime}$}

\begin{document}

\title{Illuminating the Diffuse Radio
Emission in Low-Mass Cluster: Abell 13}

\author[orcid=0009-0005-2553-6973,sname='S Anand']{Nasmi S Anand}
\affiliation{Indian Institute of Techonology Indore, Simrol, Madhya Pradesh, India, 453552.}
\email[show]{nasmisanand67@gmail.com}  

\author[orcid=0000-0001-8194-8714, sname='Chatterjee']{Swarna Chatterjee} 
\affiliation{Centre for Radio Astronomy Techniques and Technologies, Department of Physics and Electronics, Rhodes University, Makhanda 6140, South Africa}
\email{swarna.chatterjee16@gmail.com}

\author[orcid=0000-0001-8721-9897, sname='Raja']{Ramij Raja} 
\affiliation{Centre for Radio Astronomy Techniques and Technologies, Department of Physics and Electronics, Rhodes University, Makhanda 6140, South Africa}
\email{ramij.amu48@gmail.com}

\author[orcid=0000-0002-1372-6017, sname='Rahaman']{Majidul Rahaman} 
\affiliation{Institute of Astronomy, National Tsing Hua University, Hsinchu 300013, Taiwan}
\email{rmajidul@gmail.com}

\author[orcid=0000-0002-5333-1095, sname='Datta']{Abhirup Datta}
\affiliation{Indian Institute of Techonology Indore, Simrol, Madhya Pradesh, India, 453552.}
\email{abhirup.datta@iiti.ac.in}


\begin{abstract}

Recent advances in high-sensitivity radio observations have uncovered a population of faint, ultra-steep-spectrum sources in galaxy clusters, commonly known as radio phoenixes. However, their observational classification remains poorly constrained due to the limited number of confirmed detections. This study presents a detailed multi-frequency, high-sensitivity, and high-resolution analysis of diffuse radio emission in the merging galaxy cluster Abell 13. Using GMRT (147.5 MHz), uGMRT (400 MHz), ASKAP-low (887.5 MHz), and MGCLS (1284 MHz) images, we detect complex, filamentary diffuse emission with a largest linear extent of 521 kpc. This emission originates from the cluster center and extends westward, confined within the X-ray-emitting intra-cluster medium (ICM). \textit{Chandra} X-ray data confirm that Abell 13 is undergoing a merger, and the radio morphology reflects signatures of this ongoing dynamical activity. We observed filamentary structures extending towards east-northeast and southwest directions. The spectral index across the emission appears irregular and lacks a coherent spatial gradient. The integrated spectrum reveals a steep spectral index of $-1.85 \pm 0.05$ and a spectral curvature of $-0.93\pm 0.21$. These spectral properties, along with the observed morphology and brightness distribution, are consistent with a re-energization of a fossil radio plasma driven by adiabatic compression, supporting the classification of the emission as a radio phoenix.


\end{abstract}



\section{INTRODUCTION}
\defcitealias{slee2001four}{S01}
Through hierarchical structure formation, small structures have merged over time to form increasingly larger ones, eventually forming galaxy clusters ($M_{500}\sim 10^{14}-10^{15}M_\odot$). Thus, the study of galaxy clusters offers a window into the history of structure formation of the universe. The enormous amount of energy released during cluster merging causes turbulence and shocks in the intra-cluster medium (ICM), which accelerates the relativistic components (up to $\sim \gamma > 1000$ ) and amplifies the magnetic field ($\mu$ G) (\citealt{feretti1996radio, sarazin2002physics, brunetti2014cosmic}). Signatures of interaction between relativistic electrons and magnetic fields can be observed as radio emissions extending over areas from kiloparsec to megaparsec scales in clusters that are not directly related to a radio galaxy. These radio emissions in galaxy clusters are commonly classified as radio halos, mini-halos, and relics based on their morphology, physical origin, and location within the cluster environment (\citealt{van2019diffuse,kempner2003taxonomy}).\\
\textit{Radio halos} are centrally located, roughly spherical emissions that extend over volumes of approximately 1 Mpc in size, typically observed in merging galaxy clusters. Two mechanisms have been proposed to explain the origin of halos: (i) second-order Fermi acceleration, where turbulence generated in the ICM during mergers re-accelerates relativistic electrons (\citealt{brunetti2016stochastic,pinzke2017turbulence}), and (ii) hadronic models, in which secondary electrons are produced through collisions between cosmic-ray protons and thermal protons in the ICM (\citealt{dennison1980formation}).\\\textit{Mini halos} have 
smaller sizes (50 kpc - 500 kpc)  and are located in relaxed cool core
clusters, which also host a powerful radio galaxy, the brightest cluster galaxy (BCG), associated with it. They 
are found to be co-spatial with X-ray emission from ICM (\citealt{giacintucci2013new,raja2020probing}). Acceleration of particles is due to the turbulence created during a minor merger or merging within the cluster, which does not disrupt the cool core of the cluster (\citealt{van2019diffuse}). \\
\textit{Radio relics} are irregularly shaped radio sources with sizes ranging from 50 
kpc to 2 Mpc, which can be further divided into three groups: radio gischt, active galactic nuclei (AGN) relic, and radio phoenix
(\citealt{kempner2003taxonomy}). \textit{Radio gischt} (\citealt{van2019diffuse}) are large, elongated structures with linear extents reaching up to $\sim$1 Mpc, typically found on the outskirts of merging galaxy clusters. These sources trace shock fronts where particles undergo acceleration through the diffusive shock acceleration (DSA) mechanism (\citealt{botteon2016shock,bourdin2013shock,chatterjee2022unveiling}). \\
\begin{figure*}
    \includegraphics[width=0.43\textwidth]{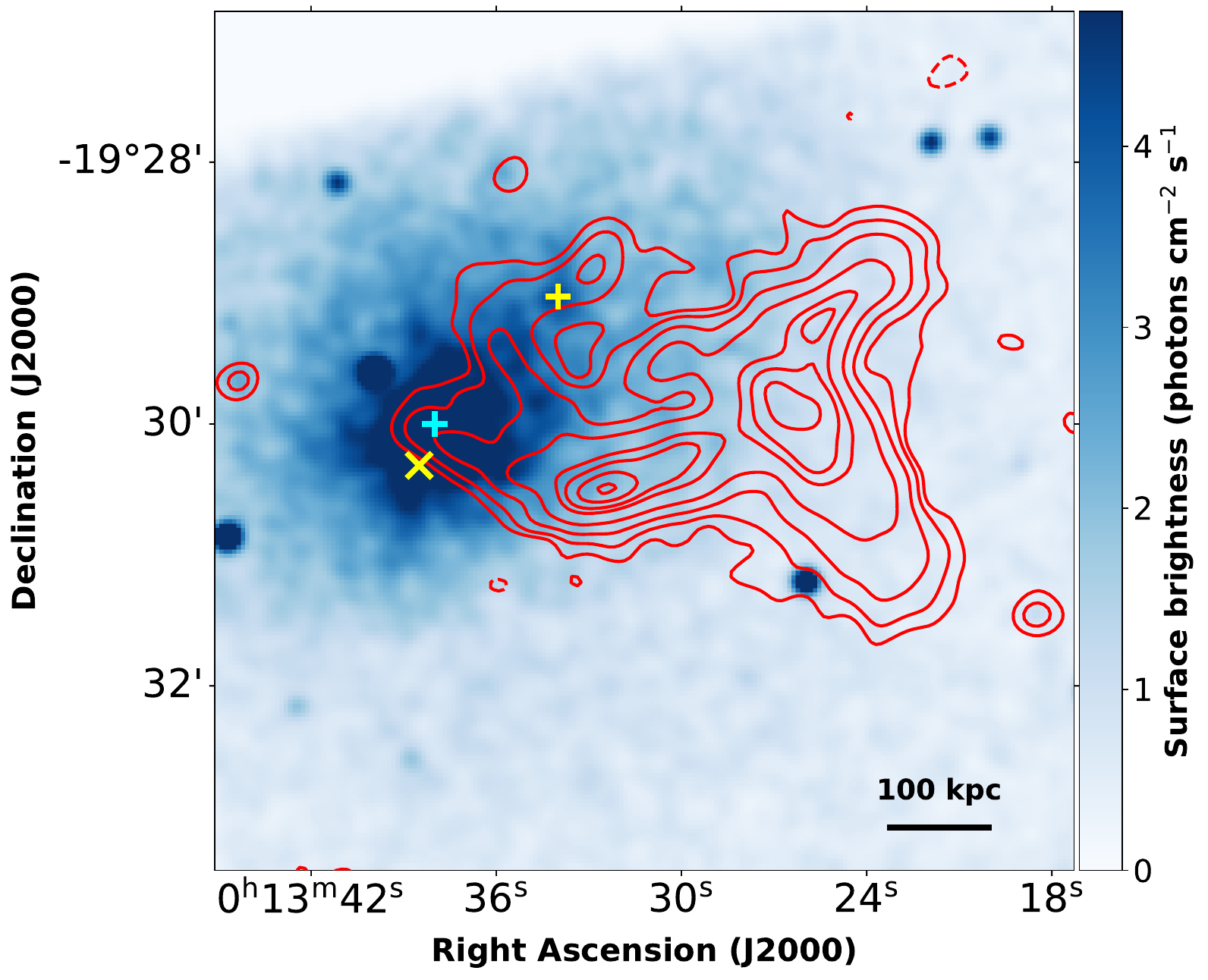}
    \hspace{0.02\textwidth}
        \includegraphics[width=0.45\textwidth]{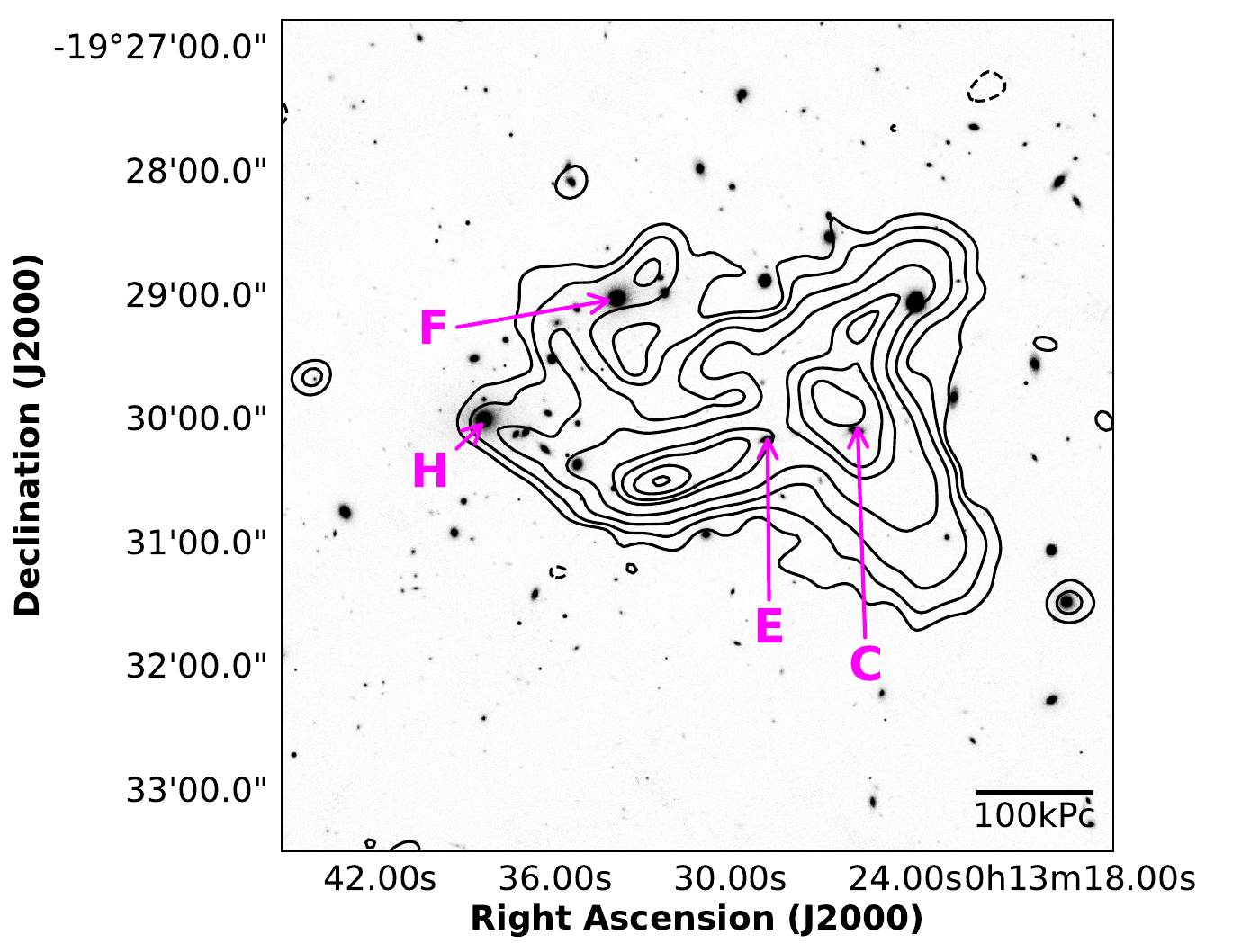}
    \caption{ \textit{Left}: Chandra X-ray image in the 0.3–8.0 keV band (pixel size $\sim$ \asec{2}), smoothed with a Gaussian kernel of $\sigma = 2$ pixels to highlight diffuse structures. The image is overlaid with uGMRT 400 MHz radio contours (red). The yellow 'X' denotes the center of the cluster, the cyan '+' is the BCG, and the yellow '+' is the position of the 2nd BCG, which is considered as the part of another sub-cluster according to \citealt{juett2008chandra}. \textit{Right}: Pan-STARRS 'r' band optical image overlaid with uGMRT 400 MHz radio contours.  The BCG is shown as 'H' and the 2nd BCG as 'F'. The labeling is done as in \citetalias{slee2001four}. Contour levels in both the images corresponds to [-3, 3, 6, 12, 24, 48, 60, 75, 90]  $\times$ $\sigma_{\mathrm{rms}}$ of image IM4 (see Table \ref{tab:complex_multicolumn_multirow}).  Negative contours are dashed. }
    \label{fig: radio and optical}
\end{figure*}
AGN relics and radio phoenixes are two classes of diffuse, steep-spectrum radio emission associated with fossil AGN plasma, and are often underexplored in high-frequency observations. The advent of high-sensitivity, low-frequency radio telescopes such as LOw-Frequency ARray (LOFAR), upgraded Giant Metrewave Radio Telescope (uGMRT), and Murchison Widefield Array (MWA) has significantly improved the detection of these faint, ultra-steep spectrum sources (\citealt{cassano2013revisiting, paul2023exploring, duchesne2021diffuse}). \textit{AGN relics} refer to the remnants of radio lobes left behind after an AGN ceases its activity. As relativistic electrons age, they lose energy through synchrotron radiation and inverse Compton (IC) scattering, causing the radio spectrum to steepen progressively until the emission becomes undetectable at radio wavelengths. However, fossil plasma may still be visible as X-ray cavities (\citealt{fujita2002chandra,mazzotta2002evidence}). \\
\textit{Radio phoenixes} are diffuse radio sources characterized by their ultra-steep, curved spectra, originating from the fossil plasma of remnant AGN lobes that have lost energy through synchrotron and inverse Compton losses. Later, these AGN lobes were re-energized by adiabatic compression, possibly triggered by shocks \citep{ensslin2001reviving}, making them observable again at sub-GHz frequencies.  Unlike radio gischt, which typically appear on the peripheries of clusters, radio phoenixes are generally located closer to the cluster center \citep{feretti2012clusters}, and are often found in low-mass clusters (\citealt{mandal2020revealing}). These sources span sizes typically in the range of $\sim$100–500 kpc and exhibit a wide variety of morphologies, from roundish \citep{kale2012spectral} to filamentary \citep{raja2023radio,bruggen2018discovery}. Due to their complex appearance, many radio phoenixes were initially misidentified as radio relics or radio gischt (\citealt{slee2001four}, hereafter \citetalias{slee2001four}). Later, multi-frequency analysis facilitates more accurate classification and characterization of the spectrum of radio phoenixes \citep{raja2023radio,randall2010radio}. Interestingly, radio phoenixes have been observed in both merging \citep{mandal2019ultra} and relaxed clusters \citep{giacintucci2020discovery}, suggesting that major cluster mergers are not essential for their formation. This is further supported by the observed anti-correlation between cluster mass ($M_{500}$, the total mass enclosed within the radius $R_{500}$, where the average density of the cluster is 500 times the critical density of the universe at that redshift) and radio power of radio phoenix, as well as between the projected distance from cluster center and the largest linear size (LLS) of the emission, suggesting that the formation of radio phoenixes is not primarily driven by the energy released during cluster mergers, but rather is linked to the state of past AGN activity (\citealt{mandal2020revealing}).\\
This paper presents the study of diffuse radio emission in the galaxy cluster Abell 13. We employed multi-frequency analysis to investigate the emission's true nature and understand the roles of AGN activity and ICM in shaping its properties. This work marks the first attempt to examine the spectral index map and spectral curvature map of this emission, using images with improved resolution and enhanced sensitivity.\\
This paper is structured as follows: in Sect. \ref{sec:A13}, we present the background and summary of previous studies on diffuse emission in Abell 13. The details of the observations and data reduction procedures are described in Sect. \ref{obs}. A comprehensive analysis of the results is provided in Sect. \ref{results}. In Sect. \ref{sec: discussion}, we discuss our findings in the context of existing theories, and finally, the conclusions of this study are summarized in Sect. \ref{sec: summary}.\\
Throughout this paper, we adopt a $\Lambda$CDM cosmology \footnote{https://docs.astropy.org/en/stable/cosmology/index.html} with $H_0 =\,70\, \mathrm{km s^{-1} Mpc^{-1}}$, $\ohm_m =\,0.3$ and
$\ohm_{\lambda} = \,0.7$. At the redshift of the cluster z = 0.099\footnote{https://ned.ipac.caltech.edu/}, \asec{1} corresponds to a physical scale of 1.83 kpc.
\begin{table}
    \centering
    \begin{tabular}{llll}
        \hline
        \rule{0pt}{1.5ex}  
        RA$_{J2000}(\alpha)$ &&& $00^h$ $13^m$ $33^s$ \\
        DEC$_{J2000}(\delta)$ &&& $-19^\circ$ \asec{28} \asec{52} \\
        Mass ($M_{500}$)$^{a}$ &&& $2.182 \times 10^{14}\,M_\odot$ \\
        Redshift (z) &&& 0.099 \\
        $L_{500}$$^{a}$ &&& $1.236  \times 10^{44}$ erg/s \\
        \hline
    \end{tabular}

    \vspace{1mm}  

    {\footnotesize
    $^{a}$~\citet{piffaretti2011mcxc}
    }

    \caption{Details of A13}
    \label{tab:A13}
\end{table}

\begin{table*}
    \begin{tabular}{|c|c|c|c|c|}
        \hline
        Array& Frequency (MHz) & Image name & Restoring beam & Sensitivity($\mathrm{Jy Beam^{-1}}$)\\ \hline
        \multirow{2}{*}{GMRT}& \multirow{2}{*}{147.5} & IM1& \asec{25}$\times$ \asec{25}, $0\degree$ & $1.07\times 10^{-3}$\\ \cline{3-5}
        & & IM2 & \asec{15} $\times$ \asec{15}, $0\degree$ & $1.05 \times 10^{-3}$ \\ \hline
        \multirow{4}{*}{uGMRT} & \multirow{4}{*}{400} & IM3 & \asec{25} $\times$ \asec{25}, $0\degree$ & $270 \times 10^{-6}$ \\ \cline{3-5}
        & & IM4 & \asec{16} $\times$ \asec{13}, $-69 \degree $ & $191 \times 10^{-6}$ \\ \cline{3-5}
        & & IM5& \asec{15} $\times$ \asec{15}, $0\degree$ & $243 \times 10^{-6}$  \\ \hline
        ASKAP & 887.5& IM6 & \asec{25} $\times$ \asec{25}, $0\degree$ & $0.25 \times 10^{-3}$\\\hline
        \multirow{3}{*}{Meerkat} & \multirow{3}{*}{1284} & IM7 & \asec{25} $\times$ \asec{25}, $0\degree$ & $24 \times 10^{-6}$ \\ \cline{3-5}
        & &IM8& \asec{15} $\times$ \asec{15}, $0\degree$ & $20 \times 10^{-6}$ \\  \hline
        
    \end{tabular}
    \caption{Details of the images }
    \label{tab:complex_multicolumn_multirow}
\end{table*}
\section{THE CLUSTER: ABELL 13 } \label{sec:A13}

\textbf{Abell 13} (hereafter A13) is a low-mass cluster located at a redshift 
of 0.099. The details of the cluster are given in the 
Table. \ref{tab:A13}. The cluster exhibits clear indications of dynamical disturbance, as evidenced by the irregular distribution of the ICM (Fig. \ref{fig: radio and optical} \textit{left}) and the presence of two distinct X-ray emission peaks associated with bright galaxies H and F (see Fig. \ref{fig: radio and optical} \textit{right} labels as per \citetalias{slee2001four}). A13 comprises two sub-clusters, A13$_ \mathrm{a}$ and A13$_\mathrm{b}$, which show a distinct separation in the redshift distribution (\citealt{fadda1996observational}). The brightest galaxies associated with these sub-clusters are galaxy H and galaxy F respectively. They occupy the very edge of the velocity distribution: galaxy F lies at the low-velocity edge of the low-velocity sub-cluster, while galaxy H resides at the high-velocity edge of the high-velocity sub-cluster. These characteristics indicate that the sub-clusters are currently undergoing a merger within cluster (\citealt{juett2008chandra}).\\
The radio diffuse emission in A13 was first reported by 
\citealt{slee1984steep}. The filamentary emission observed at 1.4 GHz using the
Very Large Array (VLA) BnA configuration, was classified as a relic by \citetalias{slee2001four}. They have proposed the possibility of galaxy C or E  being the potential source of the tailed galaxy that extends toward the southeast (see Fig. 2 of \citetalias{slee2001four}). \\
The temperature map using \textit{Chandra} observation 
(\citealt{juett2008chandra}), exposes the presence of cooler gas at the region of radio diffuse emission, with no evidence of a shock detected in the region. They have proposed two possible scenarios for the origin of this relic system : (a) the buoyant bubble scenario where the \textbf{BCG } was an active radio emitter in the past. Then, it rises from the center to a distance along with the cool gas from the cluster core. The radio lobe expands adiabatically as it rises. The relic is located at a distance of \amin{2.2} from the cluster center. Their calculation of the buoyantly rising time of the radio lobe from the BCG was consistent with the relic age calculated by \citetalias{slee2001four}, (b) from X-ray and optical analysis, it was clear that the cluster is undergoing a merger. Before merging, galaxy BCG had a cool core and AGN. During the merging, H moves towards east (as the X-ray extension is towards west), and cool core and radio lobes are stripped from the galaxy due to ram pressure from the gas in the other sub-cluster, which did not affect the stars and dark matter in BCG. They also mentioned that from \citetalias{slee2001four}, the long filament of radio emission leading back to BCG suggests that AGN remained active during the initial phase of stripping, and the X-ray signature around the galaxy says that the stripping is incomplete. However, they could not confirm the origin of this emission from existing \textit{Chandra} data. Later, a handful of low-resolution, poor sensitivity radio frequency observations were taken to understand the diffuse emission in A13 (\citealt{duchesne2021diffuse,george2017study}).  Abell 13 was first mentioned as a host of a radio phoenix by \citealt{duchesne2021diffuse}. The GaLactic and Extragalactic All-sky Murchison Widefield Array (GLEAM) survey , TIFR GMRT Sky Survey (TGSS), and Murchison Widefield Array Epoch of Reionization 0-hour field (MWA EoR project) survey could show the extended emissions that are not visible in the VLA 1.4 GHz image. An extension in the northeast direction was resolved in the TGSS image with a resolution of \asec{25}, which was absent in NVSS (\citealt{george2017study}).  \\


\section{Observations and Data Analysis } \label{obs}
In this study, we used archival data from Legacy GMRT \textbf{147.5} MHz and uGMRT Band-3 (250–500 MHz) and archival images of ASKAP-low at 887.5 MHz and MGCLS at 1284 MHz.

\subsection{uGMRT Data Reduction}\label{data reduction}
\begin{figure}
    \centering
    \includegraphics[width=0.48\textwidth]{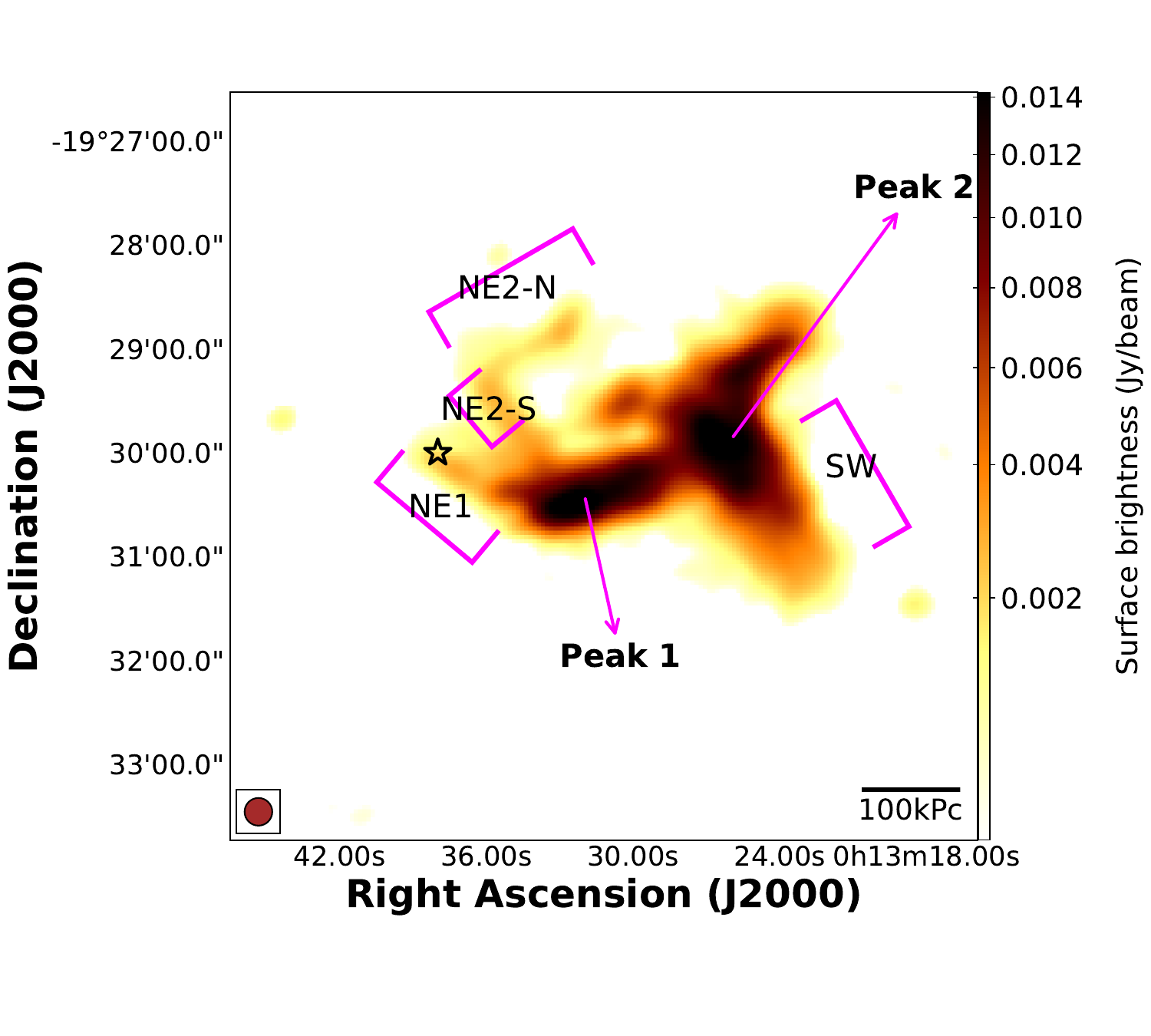}
    \caption{The uGMRT 400 MHz radio image same as in Fig. \ref{fig: radio and optical}. Contour levels are   [-3, 3, 6, 12, 24, 48, 60, 75, 90]  $\times$ $\sigma_{\mathrm{rms}}$, where $\sigma_{\mathrm{rms}} = 191\,\mathrm{\mu Jy/beam}$ of image IM4 (see Table \ref{tab:complex_multicolumn_multirow}). The restoring beam is \asec{16} $\times$ \asec{13}. The important parts of diffuse emissions labelled are discussed in Sect. \ref{results}. The black colored star marks the position of the BCG, as identified from the optical image.}
    \label{fig:A13_patch_label}
\end{figure}

For this study, we used GMRT 147.5 MHz data with a total on-source time of 513 minutes (Project Code: 20\_043) and uGMRT Band$-$3 (250$-$500 MHz) observations with a total on-source time of 161 minutes (Project Code: 32\_084). The source 3C48 was used as the flux and bandpass calibrator, and 0025$-$260 served as the phase calibrator. Data reduction for both observations was carried out using the Source Peeling and Atmospheric Modeling (SPAM; \citealt{intema2009ionospheric,intema2014spam}) pipeline. SPAM is an AIPS$-$based Python package that provides semi-automated data reduction scripts optimized for all sub-GHz GMRT observations.\\
The GMRT 147.5 MHz data, acquired using the GMRT software backend (GSB) with a bandwidth of 33 MHz, was initially pre-calibrated through standard flagging, RFI mitigation, and cross-calibration steps. Subsequently, the pre-calibrated data were processed by direction-independent and direction-dependent calibration.\\
The uGMRT Band-3 observations were recorded with both the GSB (33 MHz bandwidth) and the GMRT wideband backend (GWB; 300 MHz bandwidth). The GSB data were reduced using the aforementioned procedure. A sky model was created from the calibrated GSB data using PyBDSF (Python Blob Detector and Source Finder; \citealt{mohan2015pybdsf}). Before reducing the GWB data, it was split into six sub-bands and each sub-band was pre-calibrated using a reference frequency of 450 MHz to ensure consistent frequency averaging. The sky model derived from the GSB data was incorporated during the standard processing of the GWB data to improve image fidelity.\\
We created a multifrequency synthesis (MFS) image map using the WSClean v3.5.0 imager\footnote{https://gitlab.com/aroffringa/wsclean/-/releases/v3.5}. To bring out a better image we have used Briggs robust parameter 0 (\citealt{briggs1995high}) which compromises between RMS noise and resolution, \textit{uv}-range $<40\, \mathrm{k\lambda}$ and uvtaper 15 arcsec (image IM4 in Table. \ref{tab:complex_multicolumn_multirow}, see Fig. \ref{fig: radio and optical}, \ref{fig:A13_patch_label}). 
\section{Results}\label{results}
The diffuse emission in A13 exhibits a highly complex structure. It extends from the central BCG (black star mark in Fig. \ref{fig:A13_patch_label}) up to 521 kpc at 400 MHz toward the west-northwest. It is characterized by two distinct bright radio emissions: one directed toward the southeast (\textbf{Peak 1} in Fig. \ref{fig:A13_patch_label}) and another towards the west (\textbf{Peak 2}  in Fig. \ref{fig:A13_patch_label}). From\textbf{ Peak 2}, a faint emission further extends southwest and appears to extend beyond the detected X-ray emission region (\textbf{SW}  in Fig. \ref{fig:A13_patch_label}). Additionally, a very faint filamentary emission is observed northeast of  \textbf{Peak 1} (\textbf{NE1} in Fig. \ref{fig:A13_patch_label}), reaching the BCG, and a similar parallel filament  (\textbf{NE2-S}  in Fig. \ref{fig:A13_patch_label}) which then continues toward the northwest (\textbf{NE2-N}  in Fig. \ref{fig:A13_patch_label}). No compact radio counterparts corresponding to the optical galaxies were identified within this region, and no embedded point sources were detected, even in high-resolution imaging.\\
We have made multifrequency images to investigate emission properties in a wide
frequency range. For this purpose, we generated images with the same restoring beams to better suit the requirements of our analysis. For the GMRT 147.5 MHz and uGMRT 400 MHz images, we used the  \textit{uv}-range below 12.5 k$\lambda$  and applied a \asec{20} gaussian taper. For the multi-frequency analysis, images were restored with a fixed Gaussian beam of \asec{25} (Fig. \ref{fig:multifreq}), while a \asec{15} beam was used for the high resolution overview of emission  (Fig. \ref{fig: mgcls and ugmrt}) using WSClean parameter $-beam-shape$. Since we only have archival images of MGCLS and ASKAP$-$low, a study using different \textit{uv}$-$ranges was not possible. Therefore, we smoothed the MGCLS 1284 MHz image in the image plane from \asec{15} to \asec{25} using CASA task \textit{imsmooth}. However, when \textit{uv}-coverage differs, image-plane smoothing cannot fully recover extended emission and may introduce biases. Consequently, we cautiously interpreted the integrated spectrum, spectral index map, and spectral curvature map derived from these datasets. Similar analyses combining archival images with uGMRT data have also been carried out in other works (e.g., \citealt{chatterjee2024deciphering,pal2025ugmrt}). 

\subsection{Multifrequency Morphological comparison}\label{multifreq}
\begin{figure*}
    \centering
    \includegraphics[width=0.45\textwidth]{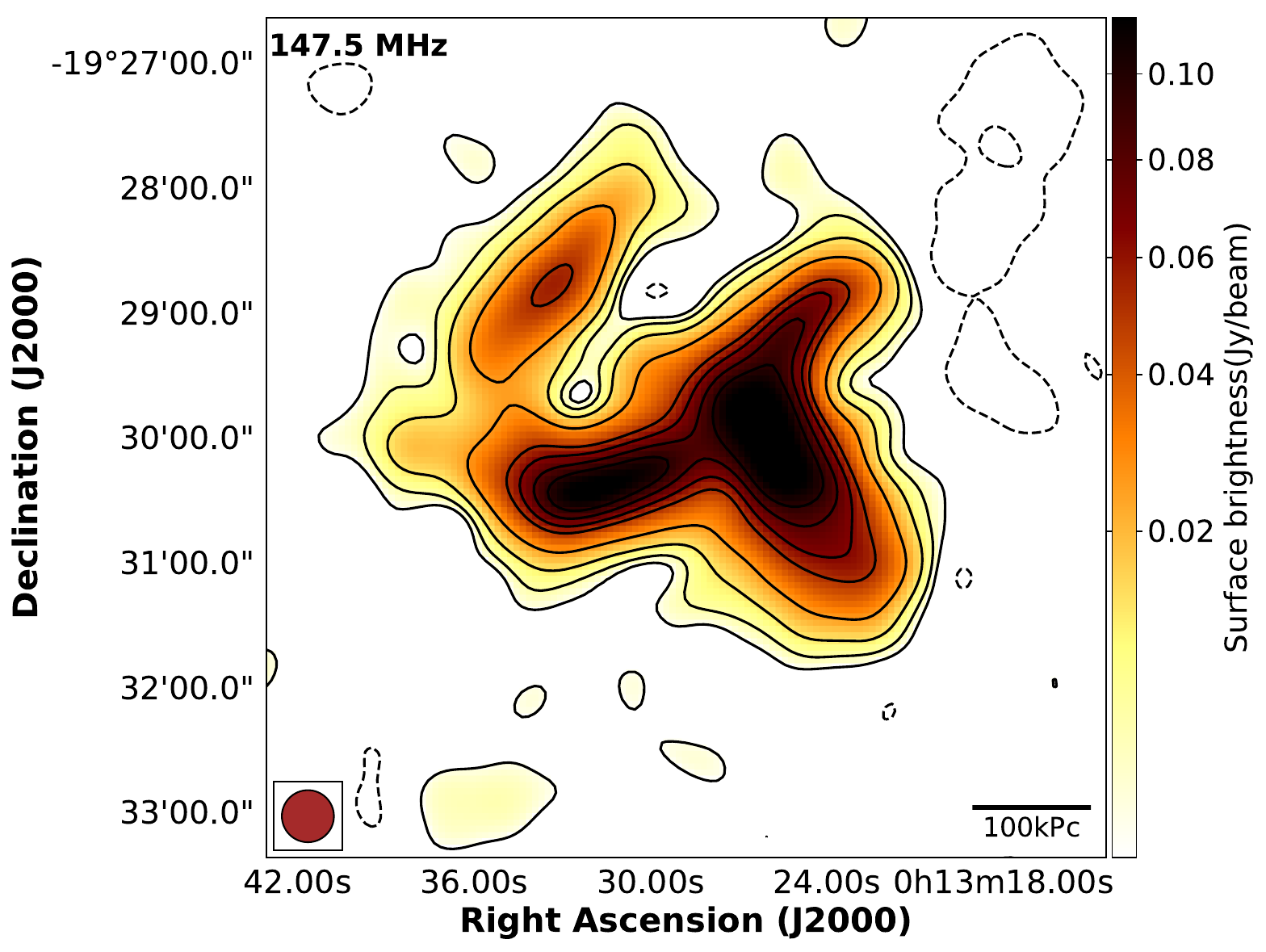}
    \hspace{0.02\textwidth}
        \includegraphics[width=0.45\textwidth]{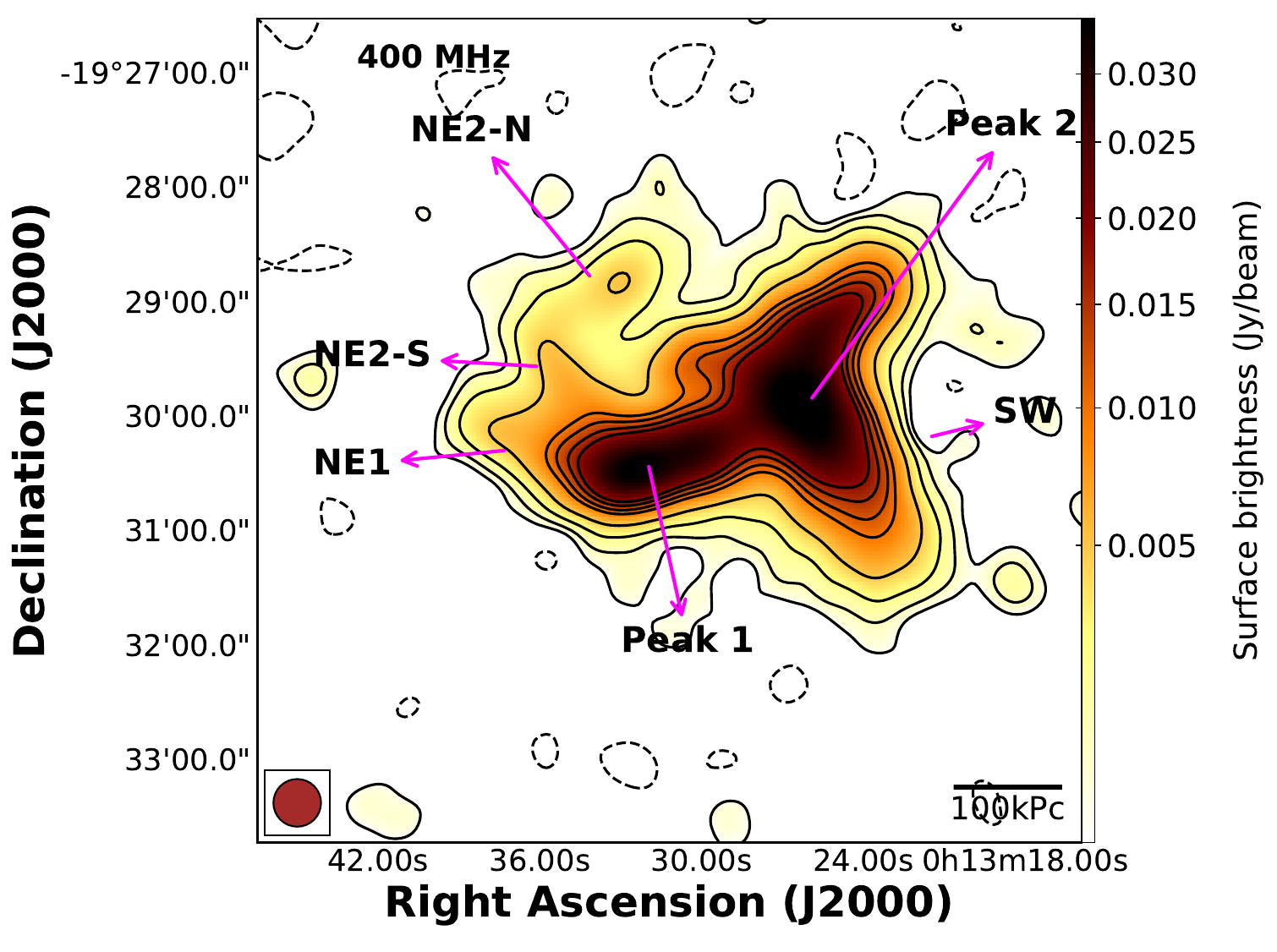}
    \vspace{1em}  
    \includegraphics[width=0.45\textwidth]{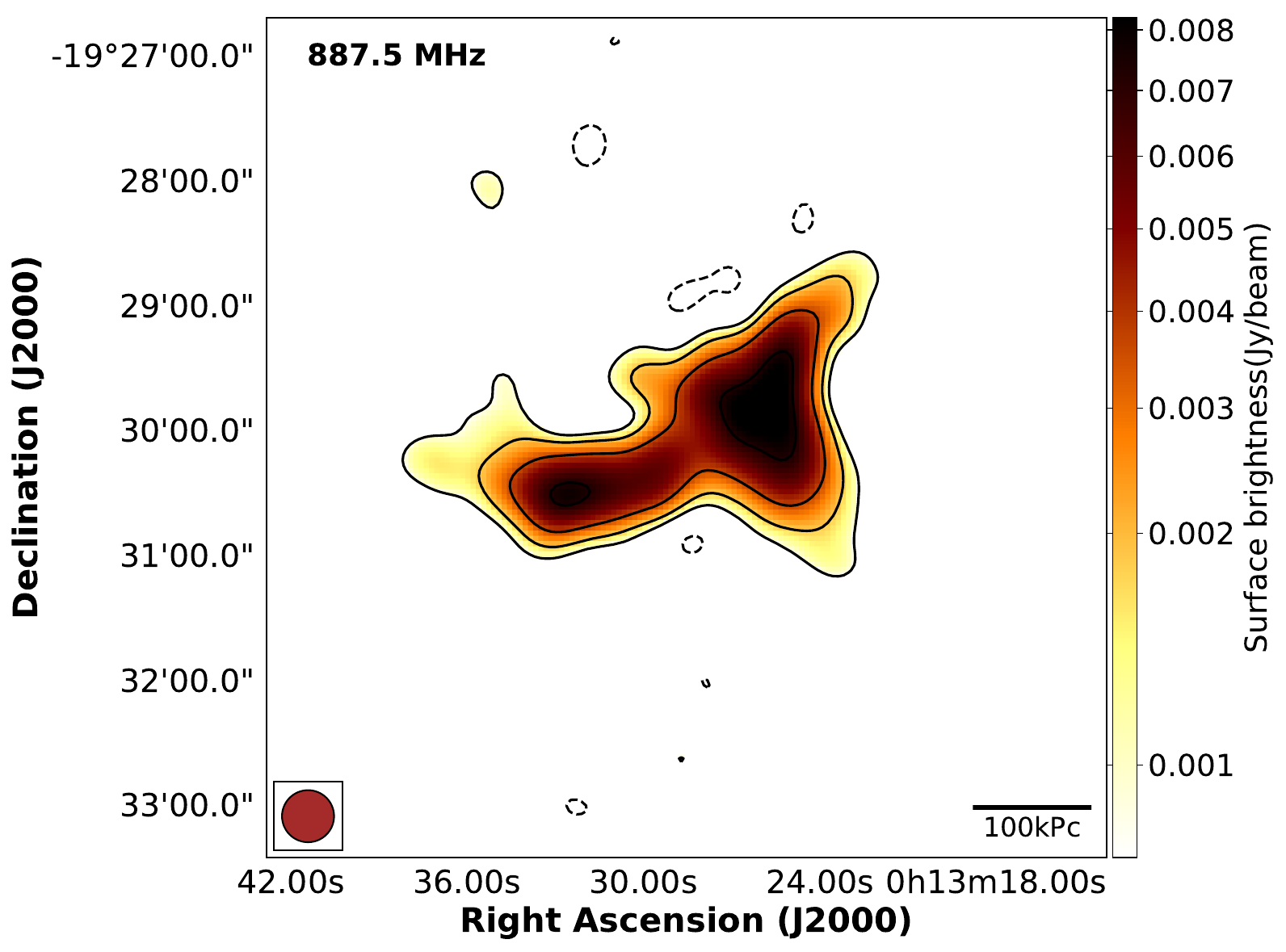}
    \hspace{0.02\textwidth}
    \includegraphics[width=0.45\textwidth]{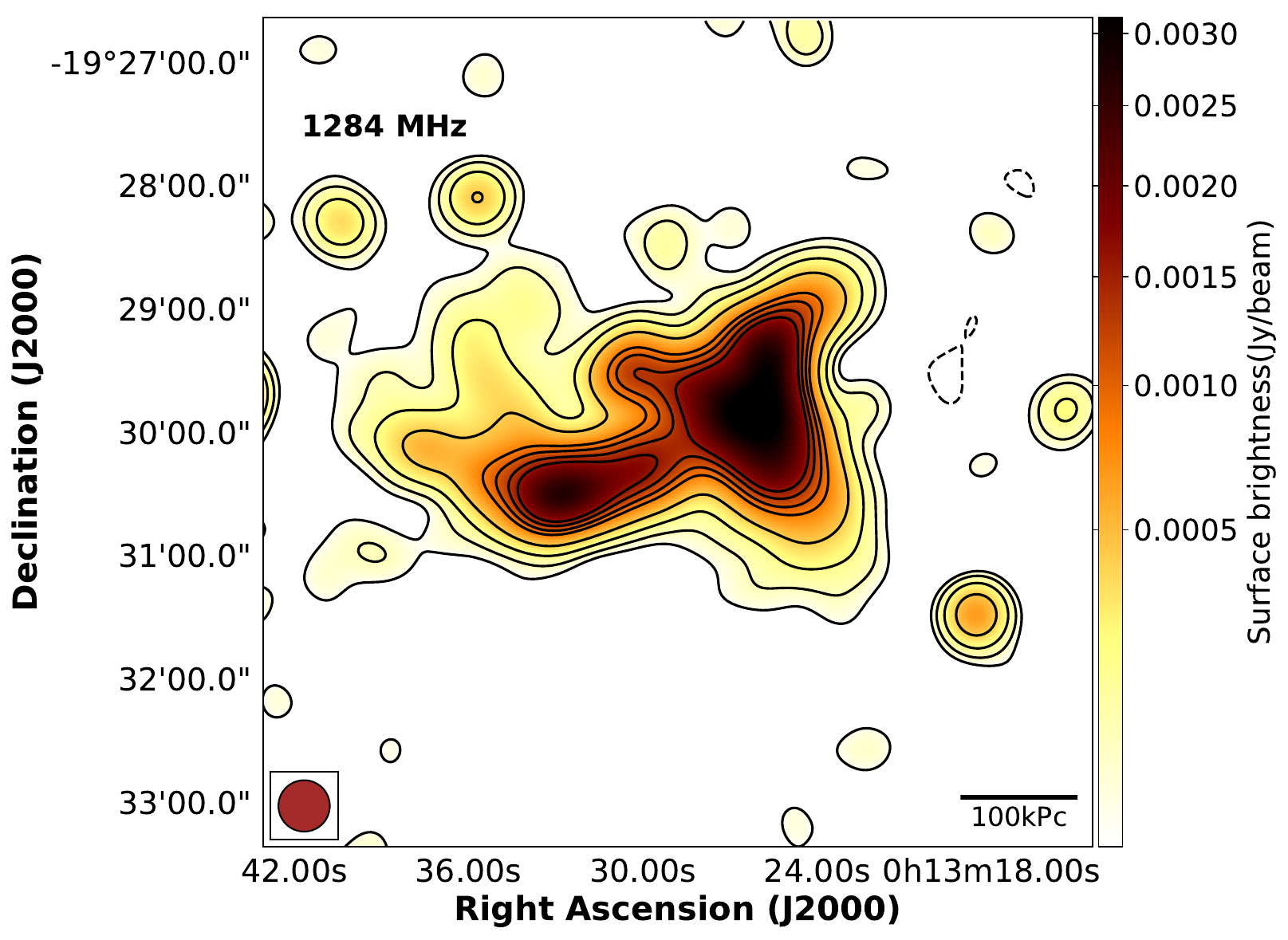}
    \caption{The radio images at frequencies 147.5 MHz (GMRT), 400 MHz 
    (uGMRT), 887.5 MHz (ASKAP-low), and 1284 MHz (MGCLS) with contour 
    levels of [-3, 3, 6, 12, 24, 48, 50, 75, 90] $\times$ $\sigma_{\mathrm{rms}}$.  The restoring beam of all images are \asec{25} $\times$ \asec{25}.  Negative contours are dashed. The left bottom  corner of every image shows  the restoring beam. The details 
    of image is given in  Table \ref{tab:complex_multicolumn_multirow} IM1, IM3, IM6, and IM7.}
    \label{fig:multifreq}
\end{figure*}
\begin{figure*}
    \centering
    \includegraphics[width=0.45\textwidth]{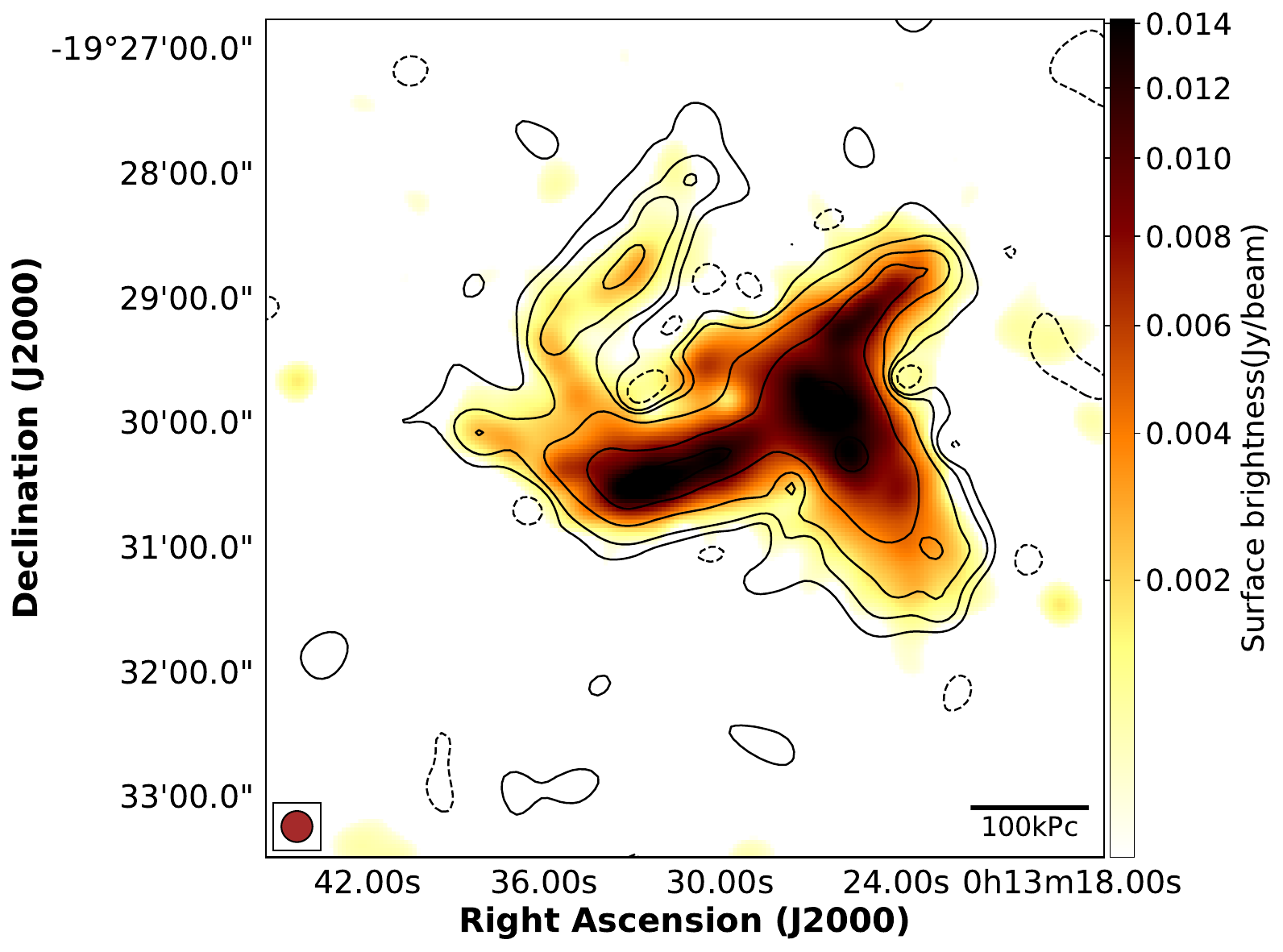}
    \hspace{0.02\textwidth}
    \includegraphics[width=0.45\textwidth]{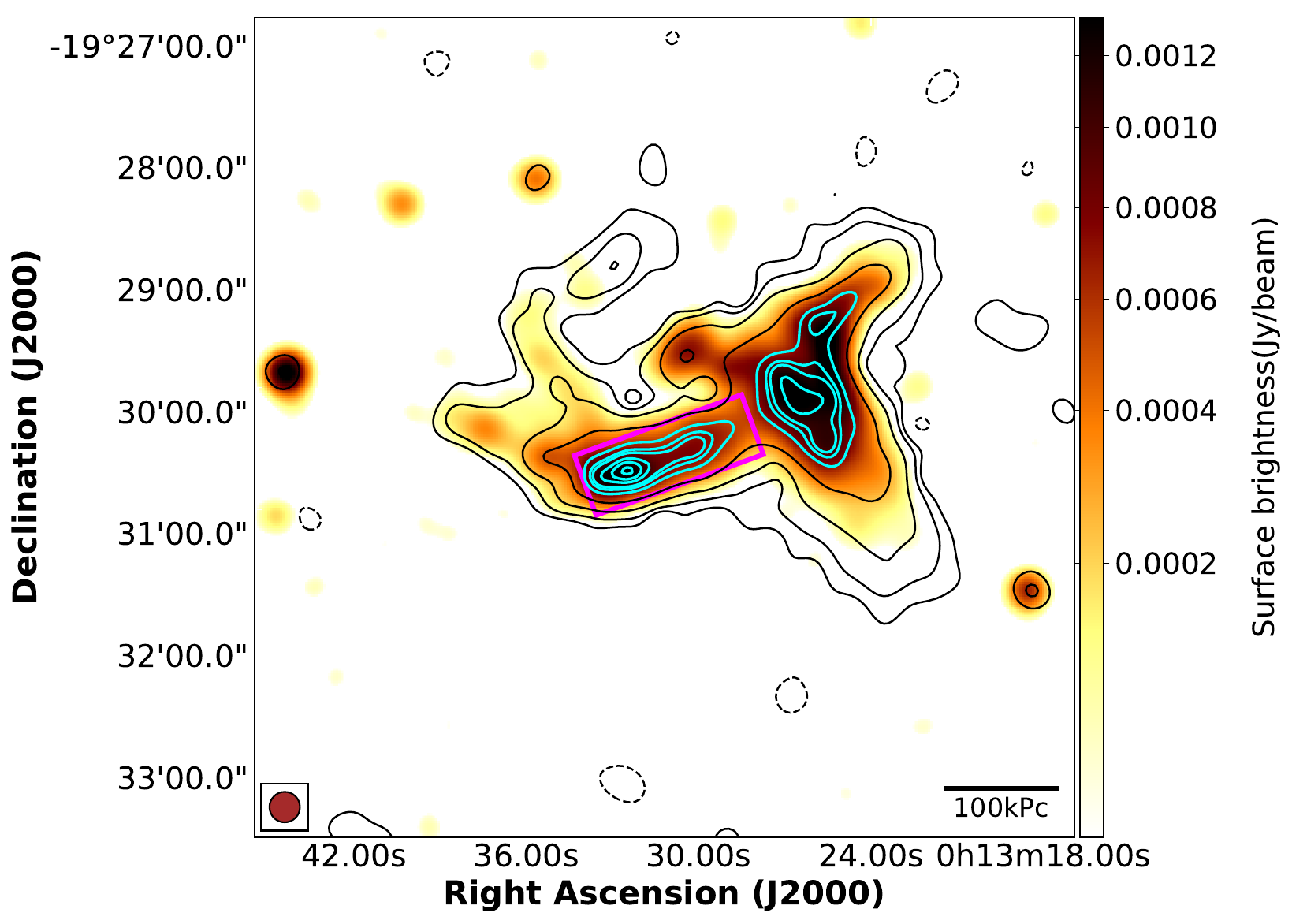}
    \caption{\textit{Left}: uGMRT 400 MHz radio image overlaid with GMRT 147.5 MHz contours. The contour levels of [-3, 3, 6, 12, 24, 48] $\times$ $\sigma_{\mathrm{rms}}$, where $\sigma_{\mathrm{rms}}$=  $1.05 \, \mathrm{mJy/Beam}$. \textit{Right}: MGCLS 1.2 GHz radio image overlaid with uGMRT 400 MHz contours. The 
    contour levels of [-3, 3, 6, 12, 24, 48] $\times$ $\sigma_{\mathrm{rms}}$, 
    where $\sigma_{\mathrm{rms}}$=  $243 \,\mathrm{\mu Jy/Beam}$. Negative contours are dashed. The cyan contour within the magenta rectangular box is used to present the FRI remanent corresponding to the \textbf{peak 1} in Fig. \ref{fig:A13_patch_label}. The images are corresponds to IM2, IM5 and IM8 in Table. \ref{tab:complex_multicolumn_multirow}}
    \label{fig: mgcls and ugmrt}
\end{figure*}

To understand the behavior of diffuse emission, we have analyzed its morphology across different
frequencies. 
Fig. \ref{fig:multifreq} allows for a side-by-side comparison of emission at various frequencies.\\

\begin{figure*}[!ht]
    \centering
    \includegraphics[width=0.4\textwidth]{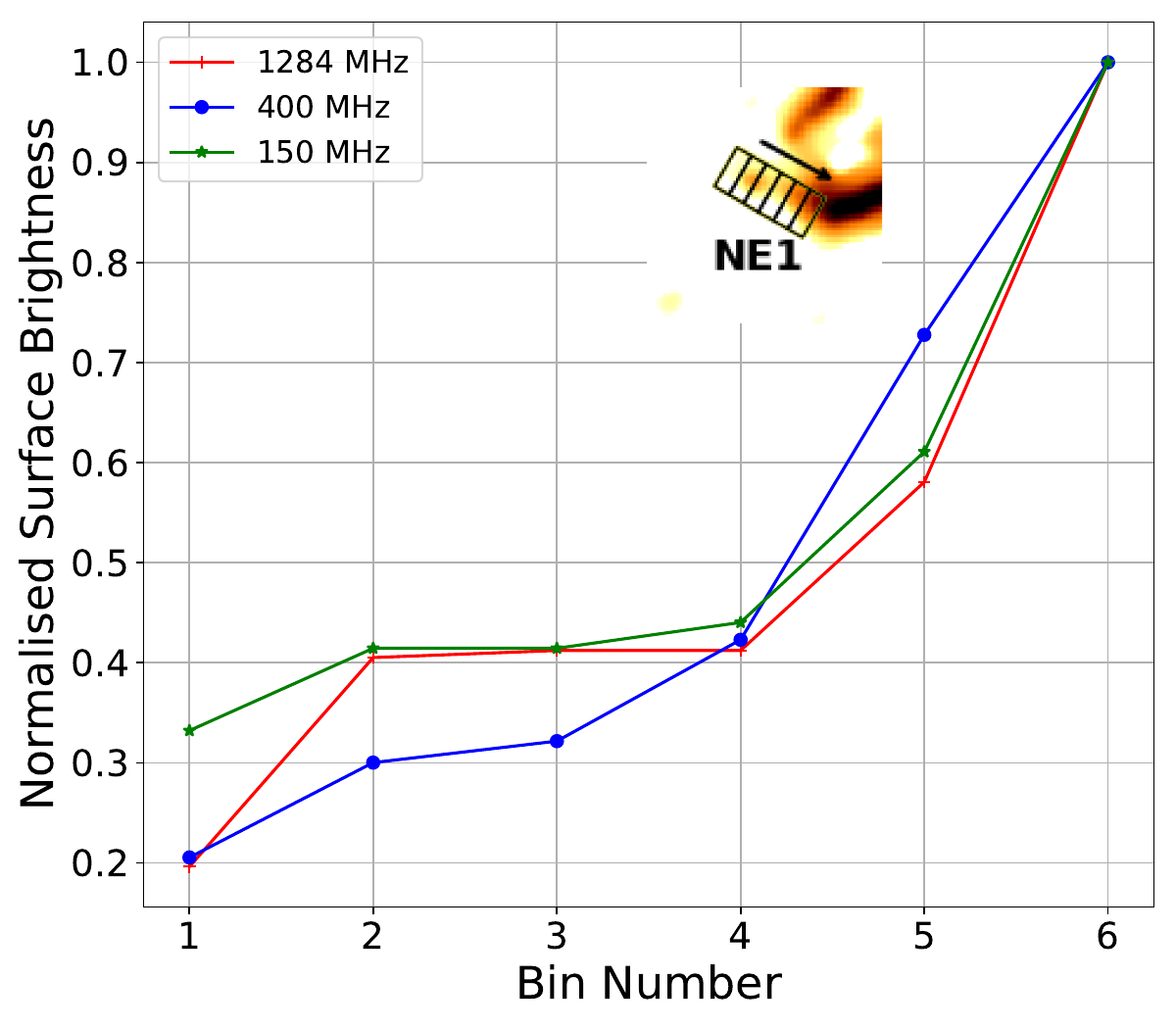}
    \hspace{0.02\textwidth}
    \includegraphics[width=0.4\textwidth]{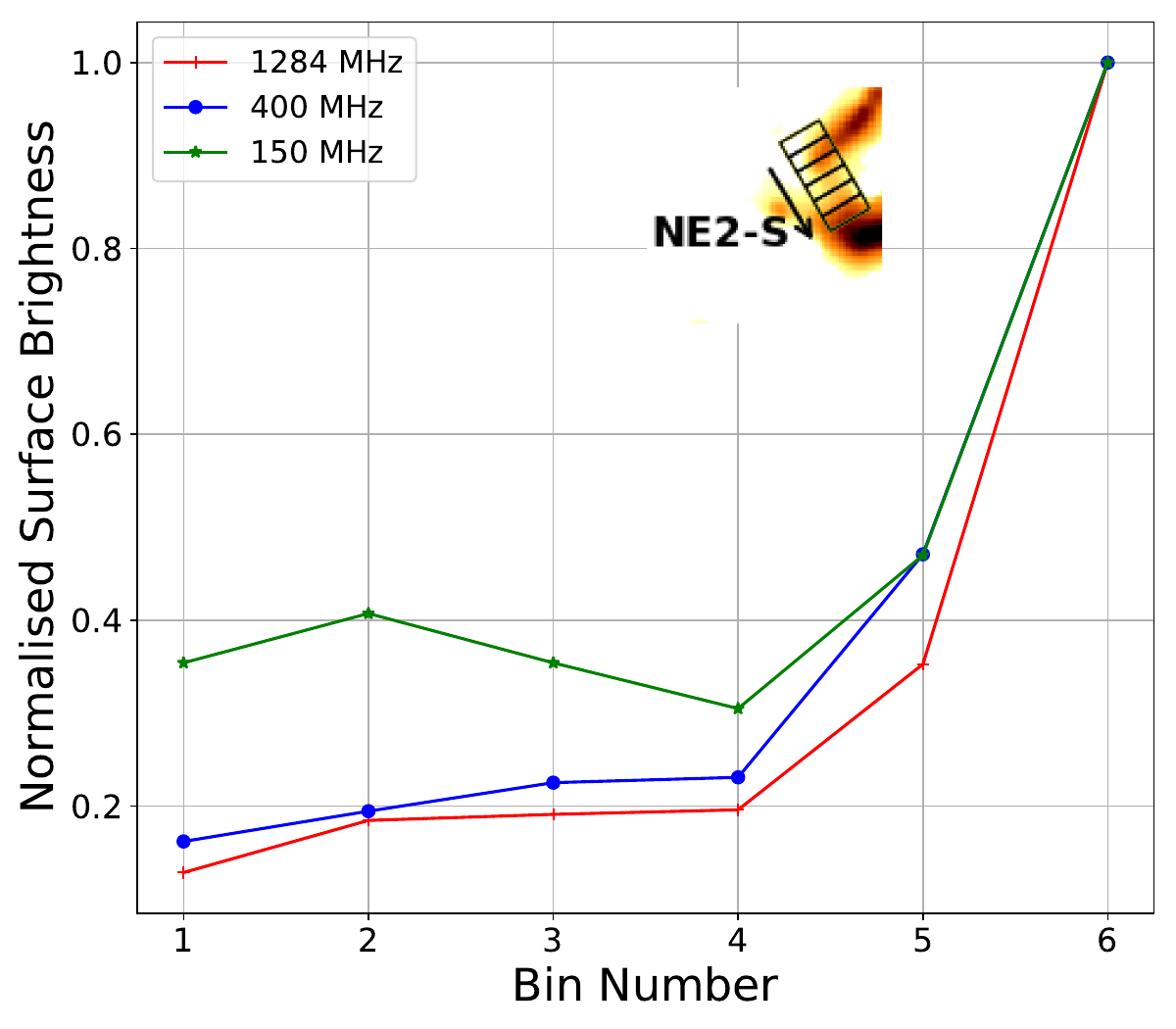}
    \caption{\textit{Left}: The surface brightness profile of the filament extending to BCG along its length. \textit{Right}: The surface brightness profile of the bridge connection between the northwest filament and the main body. The black arrow indicates the plot direction. The profile is the normalized peak surface brightness enclosed within each bin. The bin width is \asec{15}. The images used for the calculations are IM2, IM5, and IM8 in Table. \ref{tab:complex_multicolumn_multirow} }
    \label{fig:bridge and south}
\end{figure*}

In the northwestern (\textbf{NE2-N}) extension, we observe significant morphological and surface brightness variations across frequencies. The feature, along with surrounding faint emissions, appears as a broad structure at 147.5 MHz, narrows into a less pronounced filament at 400 MHz, and shows a sharp brightness drop at 887.5 MHz (Fig. \ref{fig:multifreq}). The 887.5 MHz and 400 MHz images have comparable sensitivity ($250\, \mathrm{\mu Jy/beam} $ and $270\, \mathrm{\mu Jy/beam}$, respectively, at a resolution of \asec{25}). The absence of emission at 887.5 MHz could be due to the higher observing frequency and its lower sensitivity. However, the filament is visible in the 1284 MHz image, which has a sensitivity of $24\, \mathrm{\mu Jy/beam}$. The surface brightness of the faintest emission in this filament is $72\, \mathrm{\mu Jy/beam}$ at 1284 MHz, suggesting that this structure undergoes significant radiative losses at higher frequencies.\\
Although the filament \textbf{NE1} extending toward the BCG shows different appearances at each frequency, its surface brightness profile follows a consistent trend across all frequencies (see Fig. \ref{fig:bridge and south} \textit{left}). Moreover, the filament \textbf{NE2-S} appears brighter at both 400 and 1284 MHz, but fainter at 147.5 MHz than the \textbf{NE2-N}. But, as shown in Fig. \ref{fig:bridge and south} \textit{right}, the filament \textbf{NE2-S} displays a gradual increase in brightness along its length, starting from the \textbf{NE2-N} and becoming more luminous as it approaches the main body of the diffuse emission at all frequencies, except in the 887.5 MHz image. The surface brightness of the bridge \textbf{NE2-S} is distributed across bins 3, 4, and 5. \\

The  \textbf{Peak 1} and \textbf{Peak 2} , also undergo significant change with frequencies. From the color bar, it is found that the maximum brightness value decreases by a factor of $\sim$ 10 as the frequency increases. 
As previously suggested by \citetalias{slee2001four}, \textbf{Peak 1} was identified as a possible Narrow-Angle-Tailed (NAT) remnant, with the host galaxies C or E (labeled in Fig. \ref{fig: radio and optical} \textit{right}). Any of the galaxies along the line of sight within the diffuse emission are not classified as active radio galaxies (\citealt{de2002sample, paturel2003hyperleda, seymour2007massive}). Among them, the BCG, H is the closest match based on redshift and luminosity distance criteria. Moreover, the surface brightness distribution and morphology of \textbf{Peak 1} emission align well with that of a Fanaroff-Riley I radio galaxy remanent (eg., Source 1 in A1033, \citealt{de2014new}). The steepening of the flux with frequency also indicates past AGN activity in this region (the spectral index and integrated spectrum are discussed in detail in Section \ref{int spec} and \ref{int spectrum}) making the BCG one of the possible host. \\
For a more detailed view of all parts of diffuse emission, we have created the image with a restoring beam \asec{15} (see Fig. \ref{fig: mgcls and ugmrt}).
As shown in Fig. \ref{fig: mgcls and ugmrt} \textit{left}, the 400 MHz uGMRT image is overlaid with 147.5 MHz GMRT contours. A comparison indicates that the trough between the \textbf{NE2-N} filament and the main body at 400 MHz does not align with the trough observed at 147.5 MHz. To determine whether this mismatch is due to astrometric errors or pixel misalignment, we examined the positions of point sources at both frequencies and analyzed the spectral index gradient of the same. However, no such issues were identified. The mean surface brightness of these faint emissions at 400 MHz (that are absent in the 147.5 MHz image) is 510 $\mathrm{\mu Jy/beam}$, then the surface brightness at 147.5 MHz should be 2 $\mathrm{mJy/beam}$ (considering $S_\mathrm{\nu} \propto \nu^{\mathrm{\alpha}}$ where $\alpha^{400}_{147.5}=\, -1.435$ see Sect. \ref{int spec}). Considering this as $3\sigma_{\mathrm{rms}}$, the required $\sigma_{\mathrm{rms}}$ of 147.5 MHz is 0.66 $\mathrm{mJy/beam}$. \\
The \textbf{NE2-N} filament emission and the \textbf{SW} extension are almost absent at 1284 MHz when compared to the 147.5 MHz and 400 MHz images. This shows the unavailability of high-energy electrons. This comparison clearly shows that the most energetic relativistic seed electrons associated with this diffuse emission are concentrated at the center, especially at the two peaks. \\ 
To understand more about the distribution of relativistic particles and acceleration mechanism, we have investigated the spectral index and spectral curvature map over the whole diffuse emission.
\subsection{Integrated flux density and spectral index estimation} \label{int spec}
The integrated flux density represents the total emission of the diffuse emission. We have calculated the integrated flux density within the 3$\sigma_{\mathrm{rms}}$ contours of the images shown in Fig. \ref{fig:multifreq} using \textit{casaviewer}. The details of the images are given in Table \ref{tab:complex_multicolumn_multirow} IM1, IM3, IM6, and IM7. The flux density values measured in this manner are \textbf{$S_{147.5}=\,2.25 \pm 0.22$}  Jy, $S_{400}=\,0.54 \pm 0.05$ Jy, $S_{887.5}=\,0.09 \pm 0.01 $ Jy and $S_{1284}=\,35.8 \pm 1.7 $ mJy. The flux density uncertainties are calculated using 
\begin{equation}
    \sigma_\mathrm{S} = \sqrt{(\sigma_{\mathrm{cal}} S)^2 + (\sigma_{\mathrm{rms}}{\sqrt{N_{\mathrm{beam}}}})^2} 
\end{equation}
where S, $\sigma_{\mathrm{cal}}$ and $\sigma_{\mathrm{rms}}$ are flux density, calibration
uncertainty and image noise, respectively. $N_{\mathrm{beam}}$ is the number of
beams present within the 3$\sigma_{\mathrm{rms}}$ contours. We assumed that the $\sigma_{\mathrm{cal}}$ corresponding to 148, 400,
887.5, and 1280 MHz to be 10\%, 10\% (\citealt{chandra2004late}), $\sim$ 15\% (\citealt{wang2022pilot})  and 5\% (e.g. \citealt{venturi2022radio}; \citealt{riseley2024meerkat}), respectively.\\
To understand the variation in flux density with frequencies, we have estimated spectral index using the equation, 
\begin{equation}
    \alpha= ln(S_1/S_2)/ln(\nu_1/\nu_2)
    \label{eq:spectralindex}
\end{equation} 
where S and $\nu$ denote flux
densities and frequencies, respectively. The error in the spectral index was calculated using the equation 
\begin{equation}
    \Delta \alpha = \frac{1}{\ln \frac{\nu_1}{\nu_2}} \sqrt{\left(\frac{\Delta S_1}{S_1}\right)^2 + \left(\frac{\Delta S_2}{S_2}\right)^2}
    \label{eq:spectraluncert}
\end{equation}
The calculated spectral index values are $\alpha^{400}_{147.5}=\, -1.43 \pm 0.14$, $\alpha^{887.5}_{400}= -2.23 \pm 0.22$ and $\alpha^{1284}_{887.5}= -2.52 \pm 0.43 $. We have found a steepening in the spectral index at high frequencies ($\alpha < -2$). To account for this steepening, a three-frequency color-color diagram between $\alpha_{\mathrm{high}}$ and $\alpha_{\mathrm{low}}$ was constructed following the approach in \citealt{kale2018study}. For this, we considered clusters hosting radio phoenixes as listed by \citealt{mandal2020revealing} and new detection of radio phoenix from (\citealt{hodgson2021ultra, pasini2022particle, rajpurohit2022deep, groeneveld2025serendipitous}), incorporating the most recent spectral index values available for those emissions. The color-color diagram is shown in Fig. \ref{fig:color-color and second order polynomial} \textit{left}. The spectral index values ($\alpha_{\mathrm{high}}$ and $\alpha_{\mathrm{low}}$) and the associated frequency information are listed in Table \ref{appendix:color}. The spectral indices for the diffuse emission from A13 were derived using the images presented in Fig. \ref{fig: mgcls and ugmrt}. The spectral index low is taken as $\alpha_{147.5}^{400}= -1.455 \pm 0.142$ and spectral index high is $\alpha_{400}^{1284}=-2.357 \pm 0.096$. The red line shows the 1:1 line (where $\alpha_{\mathrm{high}}=\alpha_{\mathrm{low}}$), and cluster A13 is located on the lower side of this line, showing spectral steepening at higher frequencies. According to \citealt{mandal2020revealing}, a large fraction of radio phoenix in the study possess spectral steepening at high frequencies.

\subsection{Integrated spectrum}\label{int spectrum}
\begin{figure*}[!ht]
    \centering
    \includegraphics[width=0.45\textwidth]{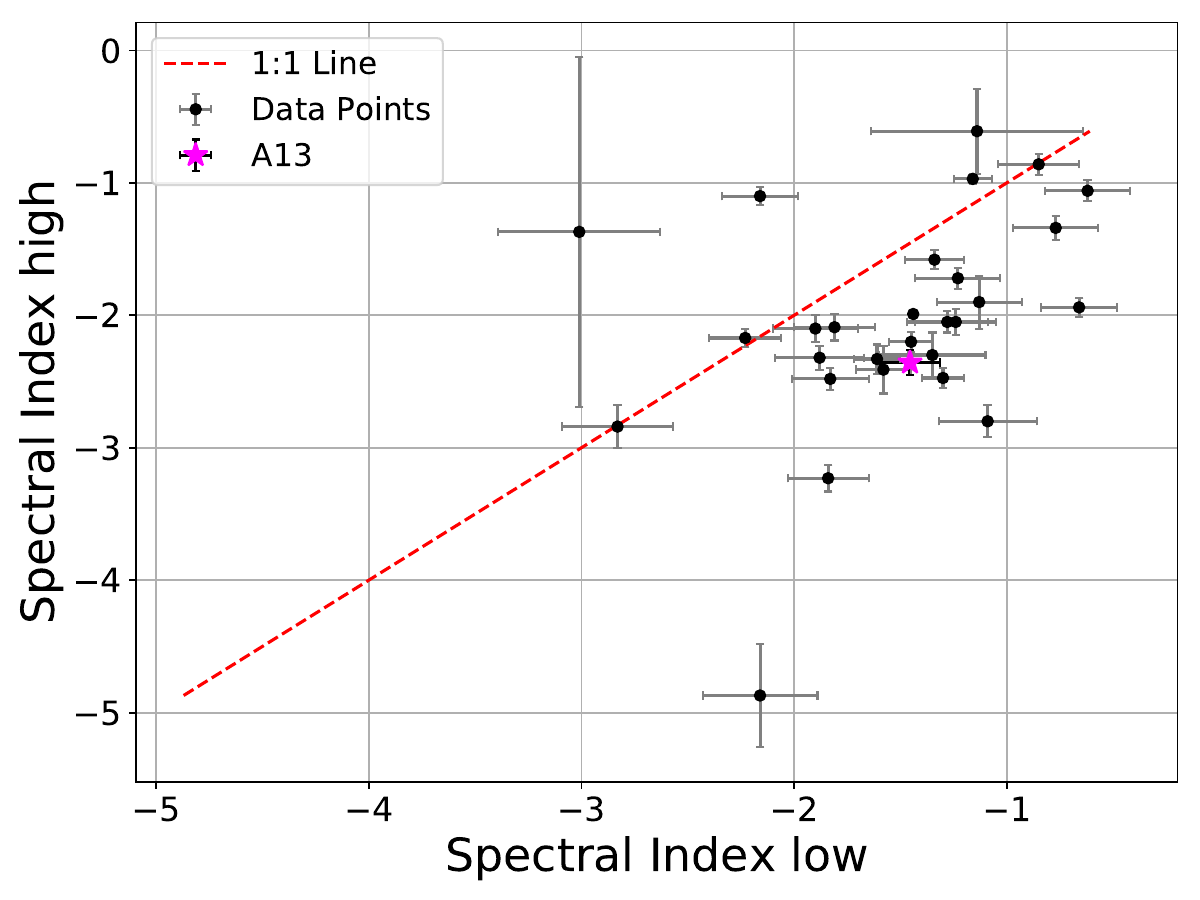}
    \hspace{0.02\textwidth}
    \includegraphics[width=0.45\textwidth]{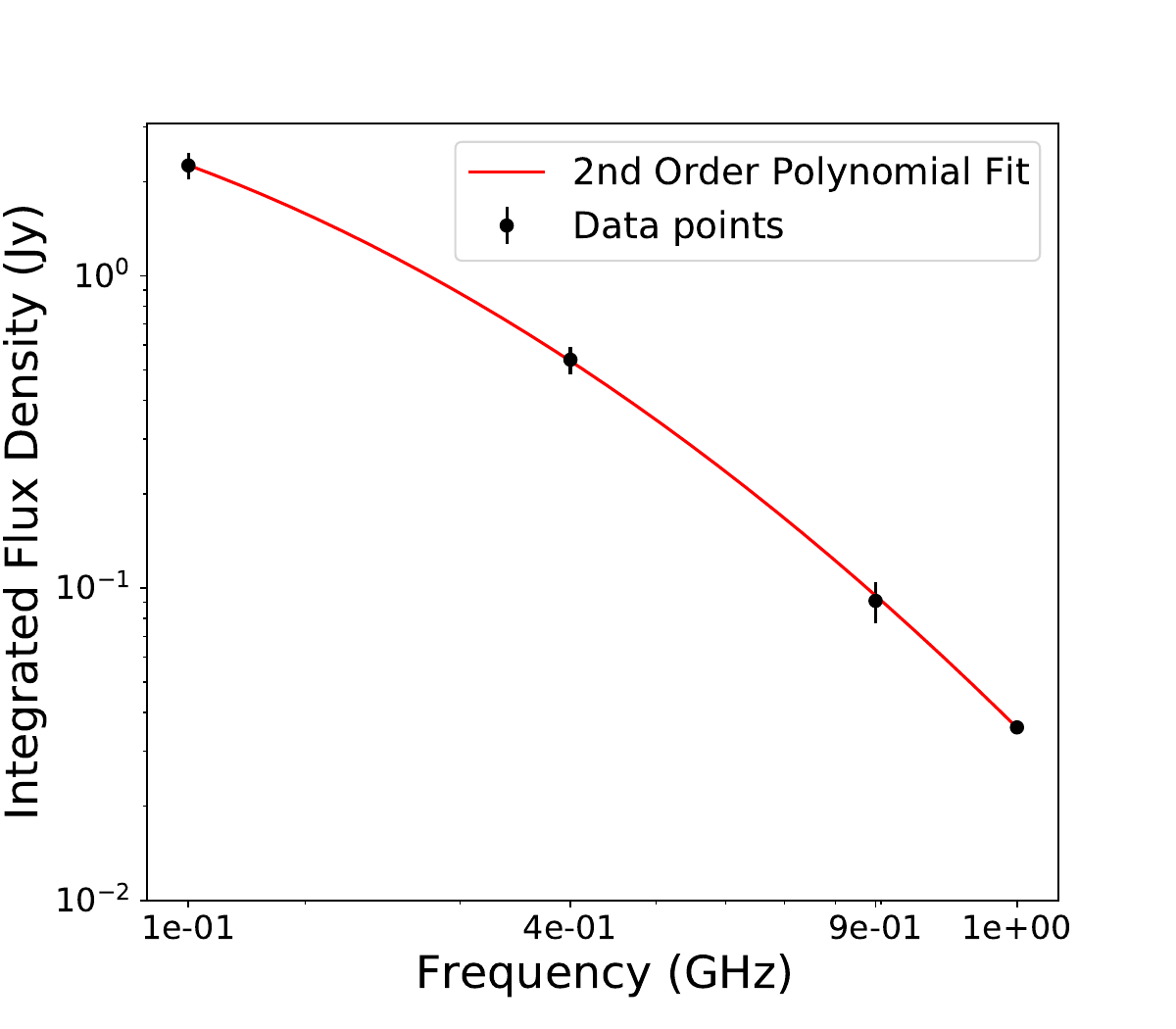}
    \caption{\textit{Left}: The color-color diagram. The spectral index values are taken from literature \citealt{bruggen2018discovery,kale2018study,mandal2019ultra,mandal2020revealing,raja2023radio,whyley2024understanding}. The spectral index values are given in Table \ref{appendix:color}. The magenta \textbf{'*'} denotes the A13, showing a steepness in high frequency. \textit{Right}: Integrated curved spectrum of emission. The best fit parameters are $a = -0.93 \pm 0.21$, $b = -2.60 \pm 0.15$ and $c=-1.15 \pm 02$. } 
    \label{fig:color-color and second order polynomial}
\end{figure*}

In Fig. \ref{fig:color-color and second order polynomial} \textit{right}, we have fitted a second order polynomial of the form $y=ax^2 + bx+c$, where $y=log(S)$, $x=log(\nu)$, $a=$spectral curvature (SC), $b=$ spectral index and c is a constant. The best-fit parameters are $a = -0.93 \pm 0.21$, $b = -2.60 \pm 0.15$ and $c=-1.15 \pm 0.02$. The spectral curvature (SC), defined as the second-order derivative of $\log(S)$ with respect to $\log(\nu)$, corresponds to the parameter $a$, thus $SC = -0.93 \pm 0.21$. The spectral index, given by the first-order derivative of $\log(S)$, is a function of frequency (\citealt{di2018deep}). At 400 MHz, the spectral index is $-1.85 \pm 0.05$, steepening to $-2.78 \pm 0.19$ at 1284 MHz. This curvature is characteristic of remnant radio galaxy lobes, where radiative and inverse Compton losses dominate \citep{murgia2011dying, brienza2016lofar}. Such ultra-steep spectra (USS, $\alpha < -2.5$) indicate aged electron populations and  inefficient particle acceleration, tracing low-energy electrons ($\gamma \sim 100$–$1000$), serving as fossil records of past AGN activity or mergers.
\subsection{Spectral Index map}
\begin{figure*}
    \centering
    \begin{minipage}{0.45\textwidth}
        \centering
        \includegraphics[width=\textwidth]{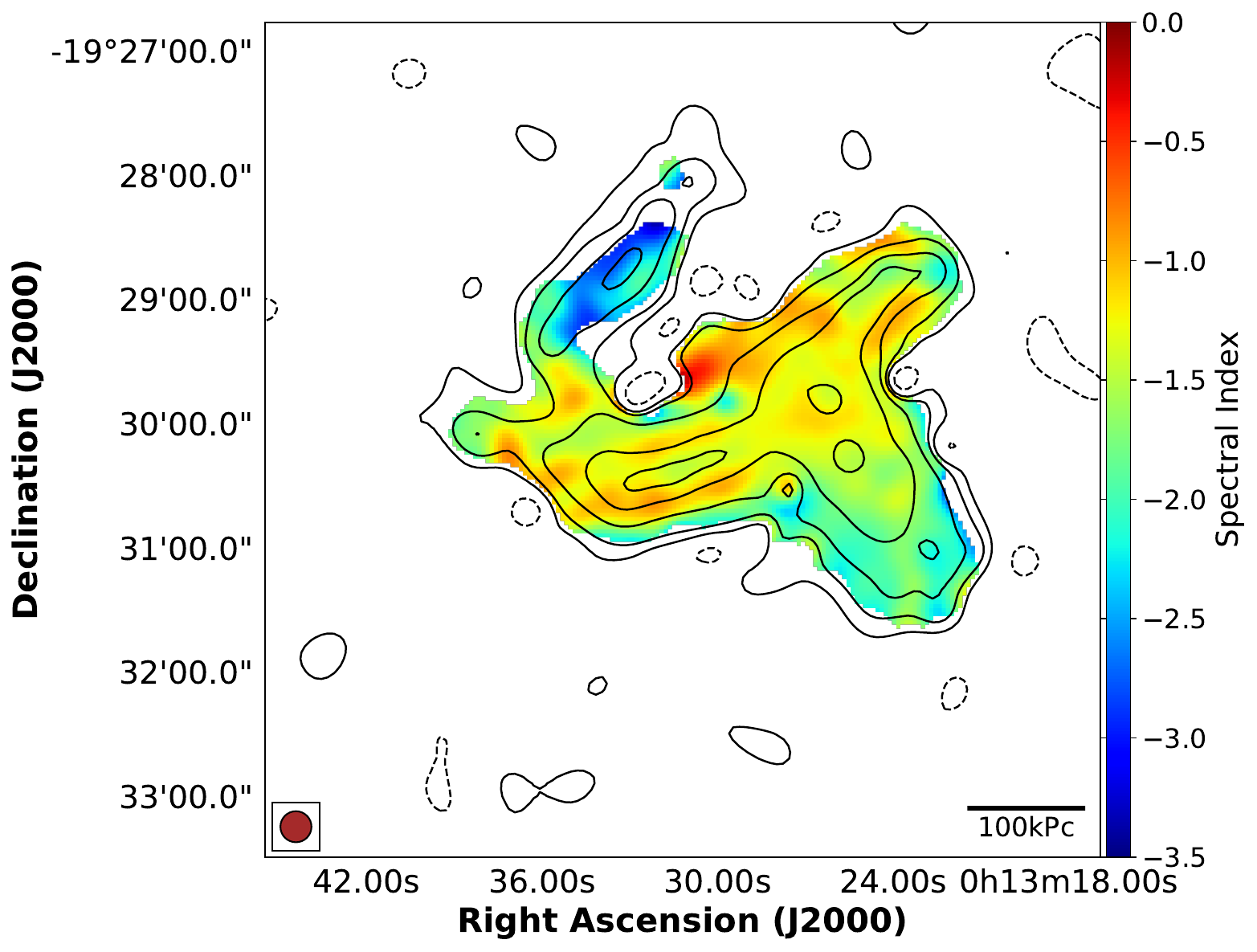}
    \end{minipage}
    \hfill
    \begin{minipage}{0.45\textwidth}
        \centering
        \includegraphics[width=\textwidth]{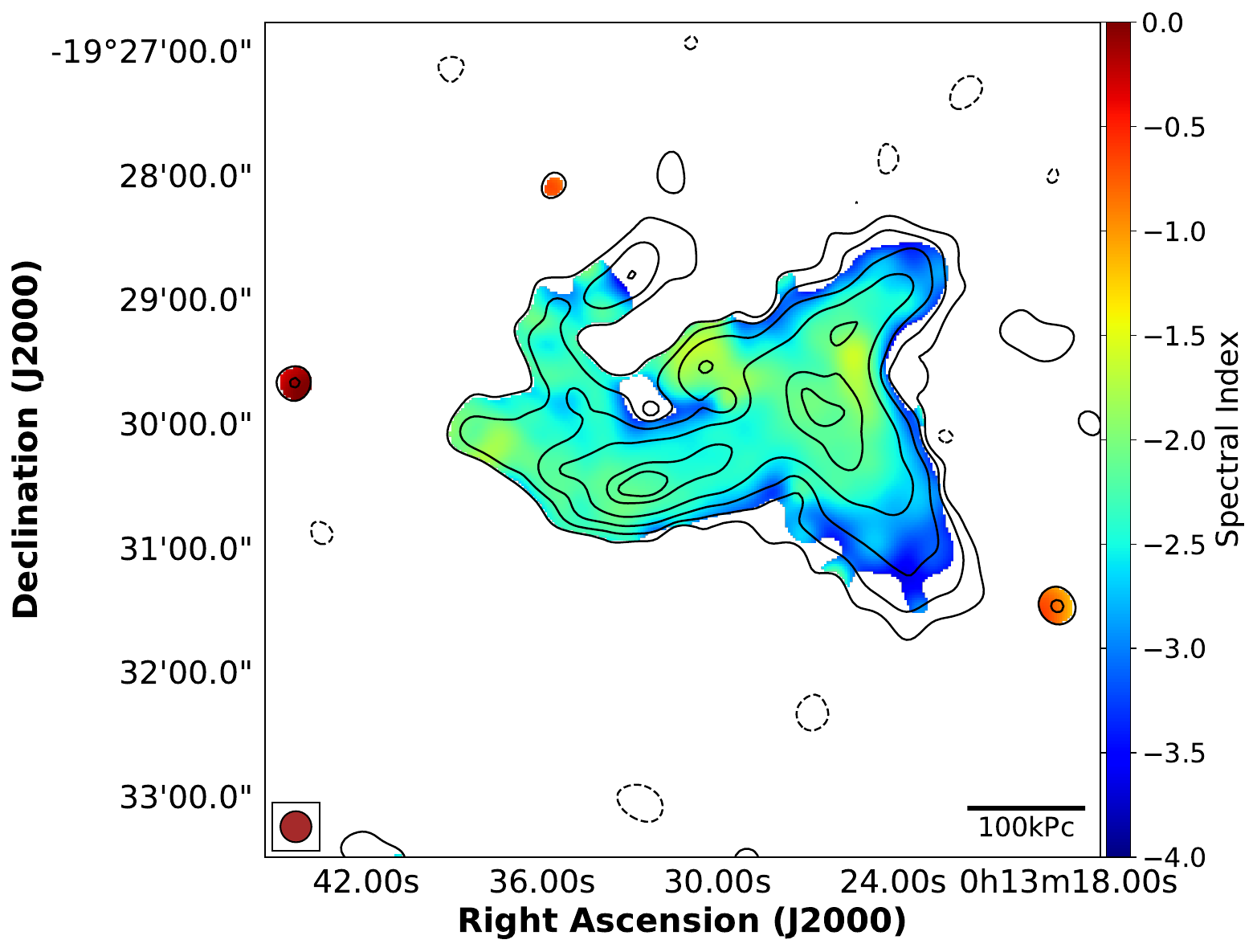}
    \end{minipage}
    \vspace{1em}
        \begin{minipage}{0.45\textwidth}
        \centering
        \includegraphics[width=\textwidth]{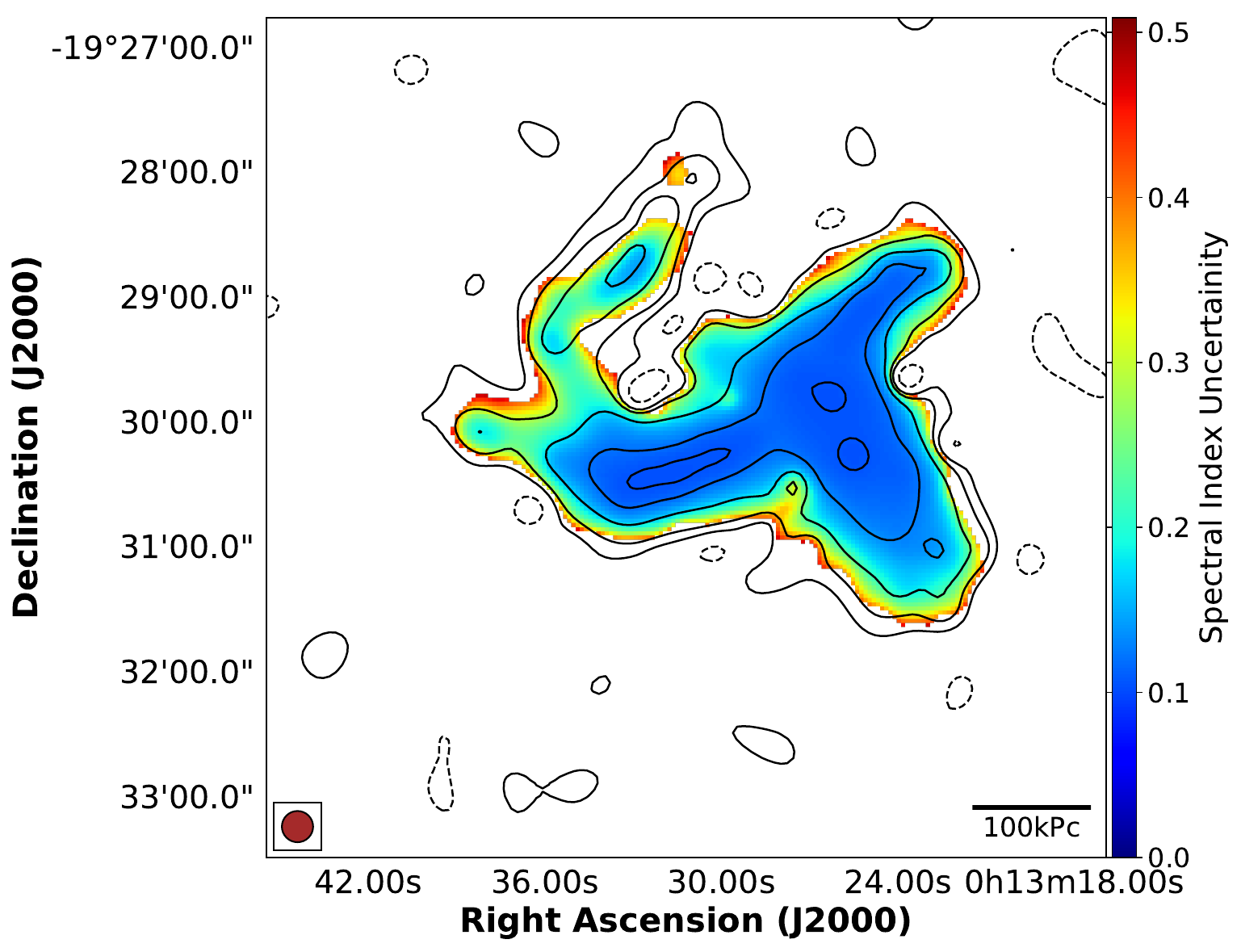}
    \end{minipage}
    \hfill
    \begin{minipage}{0.45\textwidth}
        \centering
        \includegraphics[width=\textwidth]{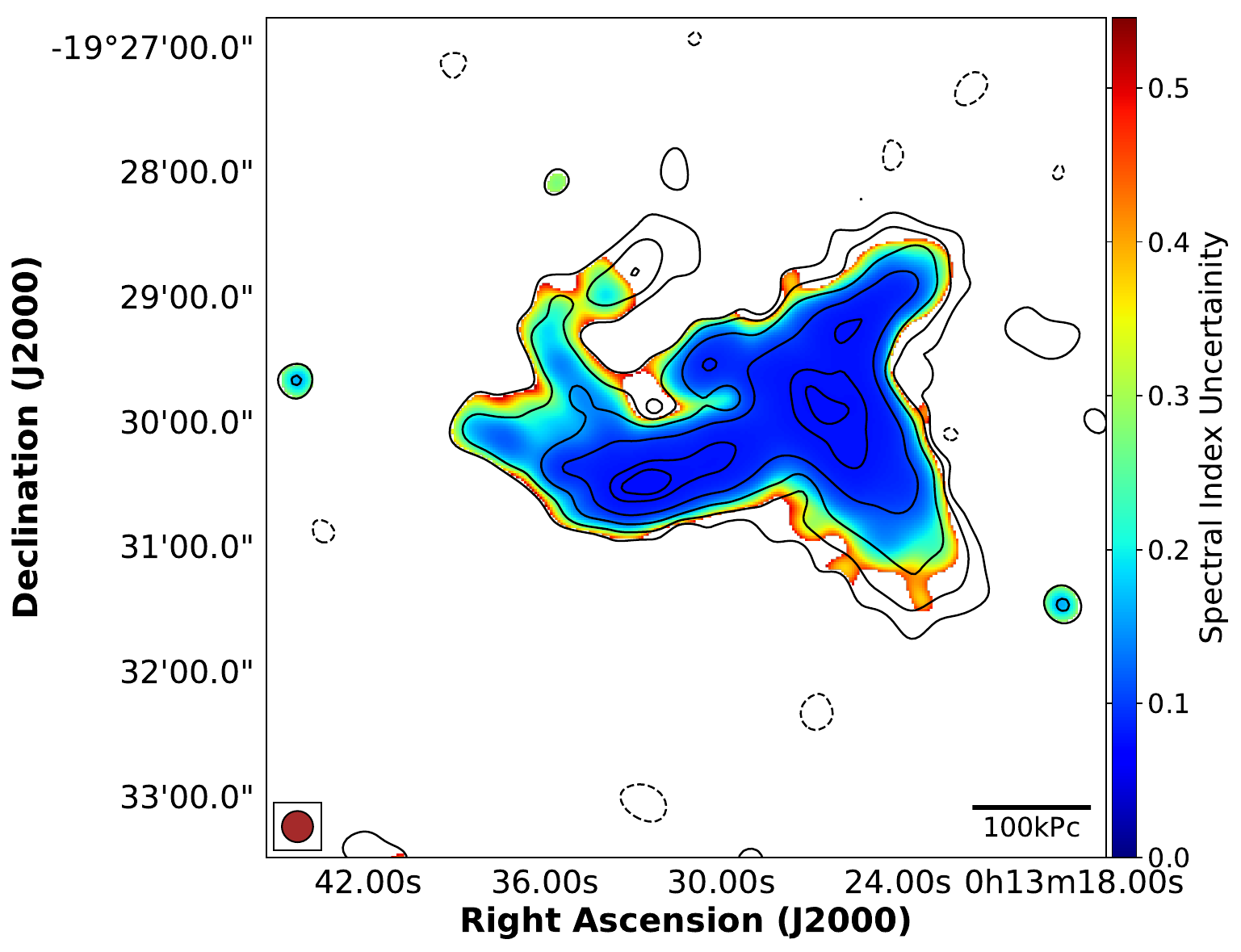}
    \end{minipage}
    \caption{Top row: spectral index maps between 400-147.5 MHz and 1284-400 MHz ( IM2, IM5, IM6  Table \ref{tab:complex_multicolumn_multirow}). The contours overlaid on the maps correspond to 147.5 MHz image (IM2) for top left, 400 MHz image (IM5) for top right  at levels [-3, 3, 6, 12, 24, 48, 50, 75, 90] $\times$ $\sigma_{\mathrm{rms}}$. Negative contours are dashed. Bottom row: Corresponding uncertainty in the spectral index.  }
    \label{fig:specmap}
\end{figure*}

The spatially resolved spectral index map is important for understanding the age of particles and acceleration process involved.
Freshly accelerated particles typically show a flatter spectral index, which steepens as they lose energy over time. The diffuse emission in cluster A13 is unique with its faintest filamentary emissions. To incorporate that emission and unveil its origin,  we have only used GMRT 147.5 MHz, uGMRT 400 MHz and MGCLS 1284 MHz images with restoring beam \asec{15} (see images in Fig. \ref{fig: mgcls and ugmrt}). We did not include the ASKAP-low 887.5 MHz image since it could not  map the \textbf{NE2-N} filament. Among these three images, 147.5 and 400 MHz images have been created using the same \textit{uv}$-$ranges as mentioned in section \ref{results}, making the spectral index variation between 147.5–400 MHz more reliable than one with the 1284 MHz image. \\
The images used for the spectral index map are IM2, IM5 and IM8 in Table. \ref{tab:complex_multicolumn_multirow}. To ensure pixel-to-pixel alignment between the two images used for spectral index mapping, we re-gridded one image onto the template of the other using the \textit{CASA} task \textit{imregrid}. Subsequently, the spectral index map was generated using the \textit{immath}.\\
In both spectral index maps (Fig. \ref{fig:specmap}), we did not observe any obvious gradient across the emission region except in the southwest direction. 
The spectral index values vary from $-1 \pm 0.1$ to $-2 \pm 0.28$ for $\alpha^{400}_{147.5}$ (Fig. \ref{fig:specmap}\textit{top left}), and from $-1.5 \pm 0.1$ to $-3 \pm 0.29$ for $\alpha^{1284}_{400}$ (Fig. \ref{fig:specmap}\textit{top right}). Although the central region appears relatively flatter compared to the edges, the steepest emission is observed in the \textbf{SW}, which lies outside the detected ICM X-ray emission region. The filamentary structure \textbf{NE2–N} shows steeper spectral indices toward its extremities; however, larger uncertainties in these regions limit a robust interpretation.
The patchy distribution of spectral indices suggests a mixture of relativistic electron populations with different ages or acceleration mechanisms.

\begin{figure*}
    \centering
\begin{minipage}{0.45\textwidth}
        \centering
        \includegraphics[width=\textwidth]{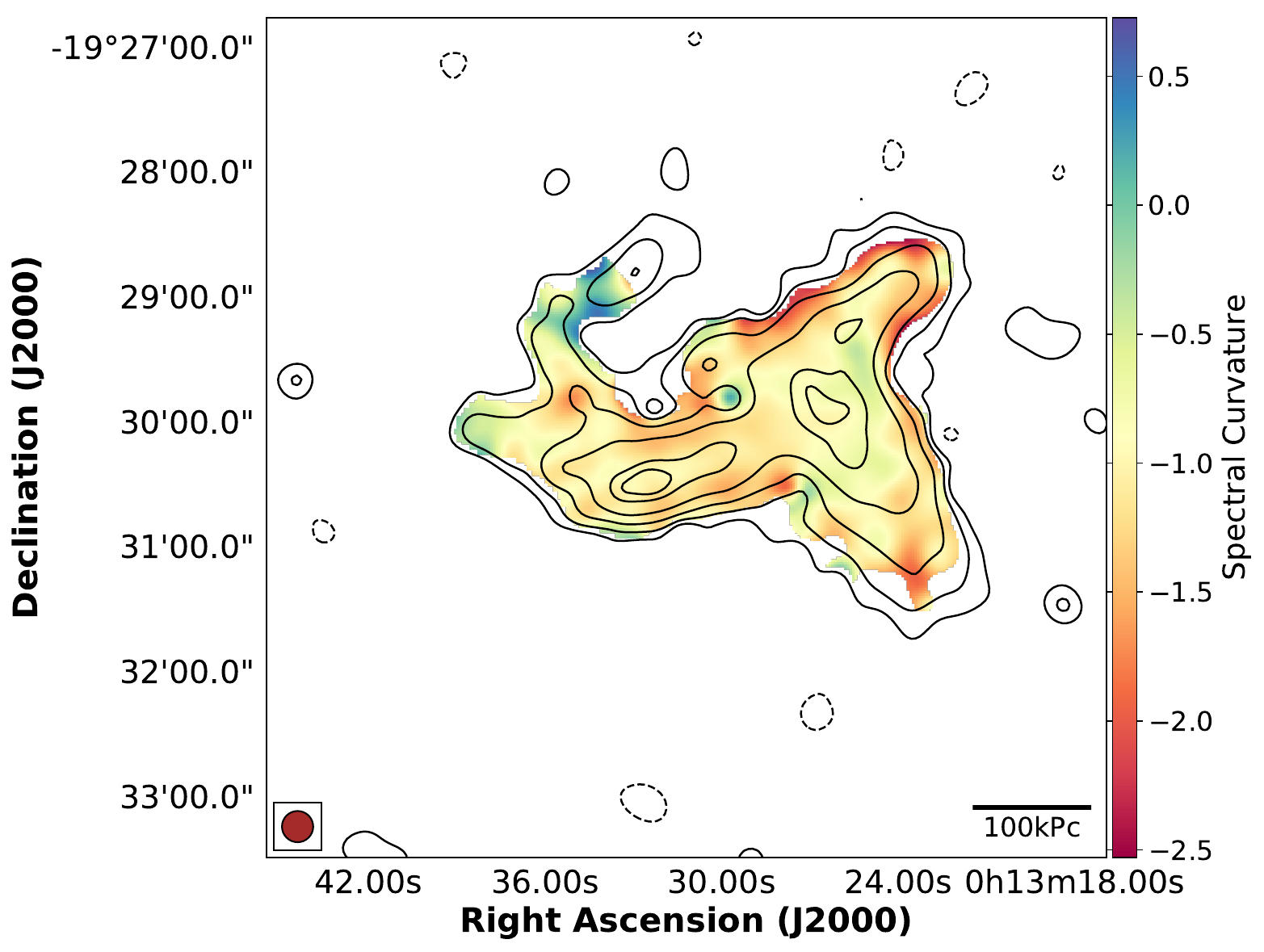}
    \end{minipage}
    \vspace{1em}
        \begin{minipage}{0.45\textwidth}
        \centering
        \includegraphics[width=\textwidth]{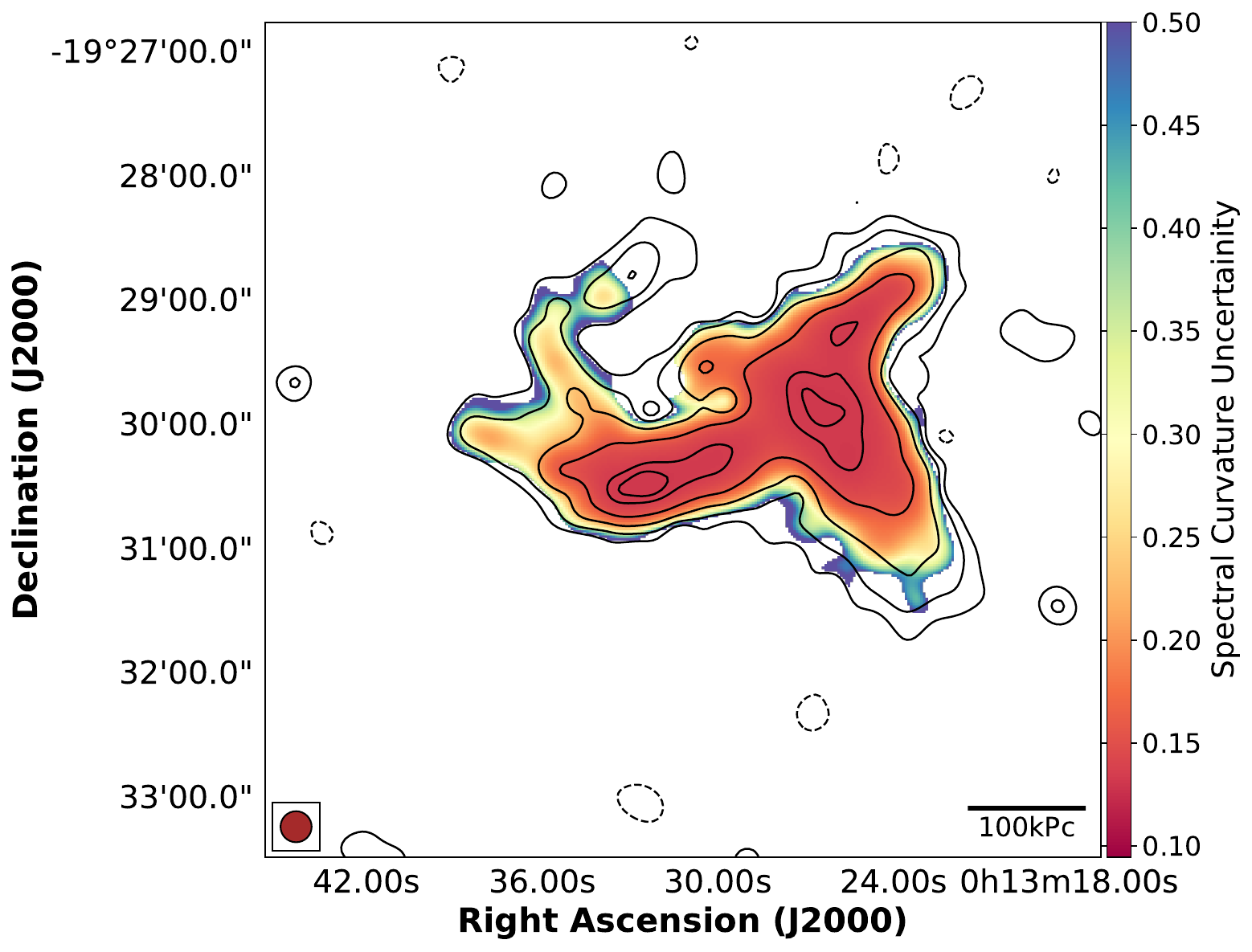}
    \end{minipage}
    \caption{ Spectral index curvature map between 1284 MHz , 400 MHz and 147.5 MHz from spectral index map showed in Fig. \ref{fig:specmap} (IM2, IM5 and   IM6 Table \ref{tab:complex_multicolumn_multirow}). The contours overlaid on the maps corresponds to 400 MHz radio image (IM5).}
    \label{fig:curv map}
\end{figure*}

\subsection{Spectral Curvature map}
We have used three of the four images in this study i.e., 1284 MHZ, 400 MHz, and 147.5 MHz to make the curvature map. Here spectral curvature map has been created using two spectral index maps of 1284 - 400 MHz and 400 - 147.5  MHz (Figures in Fig. \ref{fig:specmap} \textit{Top} row) using \textit{CASA} task \textit{immath}. The equation to derive the spectral curvature is 
 \begin{equation}
     SC= \alpha^{\mathrm{\nu_2}}_{\mathrm{\nu_3}} -\alpha^{\mathrm{\nu_1}}_{\mathrm{\nu_2}}
 \end{equation}
where, $\nu_3 = 1284 $ MHz, $\nu_2 = 400 $ MHz and $\nu_1 = 147.5 $ MHz and the spectral index values are from corresponding spectral index map. 
 The associated spectral curvature uncertainty is derived using the equation \ref{eq:curvature} (for derivation see Appendix of \citealt{raja2023radio})
\begin{equation}
    \Delta SC = \frac{1}{d} \sqrt{\left(a \frac{\Delta x}{x}\right)^2 + \left(b \frac{\Delta y}{y}\right)^2 + \left(c \frac{\Delta z}{z}\right)^2}
    \label{eq:curvature}
\end{equation}
where \\
\[
x = S_1, \quad y = S_2, \quad z = S_3
\]
\[
a = -\ln\left(\frac{\nu_2}{\nu_3}\right), \quad b = \ln\left(\frac{\nu_1}{\nu_3}\right)
\]
\[\quad c = -\ln\left(\frac{\nu_1}{\nu_2}\right), \quad d = \ln\left(\frac{\nu_1}{\nu_2}\right)\ln\left(\frac{\nu_2}{\nu_3}\right)\]
The resultant spectral curvature map and spectral curvature uncertainty map are presented in Fig. \ref{fig:curv map}.
Overall, the curvature is predominantly negative, typically between –0.5 and –1.5, with about 0.0–0.2 uncertainties. A near-zero curvature is seen at the tip of the \textbf{NE2–N} filament. However, this feature coincides with relatively significant uncertainties of around $ 0.45–0.5$. These large uncertainties render its physical interpretation ambiguous.

\section{Discussion}\label{sec: discussion}
In this study, we performed analysis of diffuse emission in the A13 at four frequencies of 147.5 MHz, 400 MHz, 887.5 MHz and 1284 MHz. The resolution and RMS noise of each image shown in this work are included in Table. \ref{tab:complex_multicolumn_multirow}. \\
We report the detection of a complex structured diffuse radio emission in Abell 13, classified as a radio phoenix \citep{duchesne2021diffuse}. The emission is located near the central region of the cluster, with a noticeable offset from the peak of the X-ray emission region of  ICM, as seen in Fig. \ref{fig: radio and optical} \textit{left}. Originating from the BCG (Fig. \ref{fig: radio and optical} \textit{right}), the radio emission extends westward, reaching an LLS of approximately 521 kpc at 400 MHz. Given that the $R_{500}$ of A13 is 0.88 Mpc \citep{piffaretti2011mcxc}, the entire radio emission lies well within the $R_{500}$ radius of the cluster. This spatial association aligns with a typical characteristic of radio phoenixes, which are often found in low-mass clusters (\citealt{mandal2020revealing}). In such low-mass clusters, the shallower gravitational potential well at the center facilitates the retention and accumulation of cosmic-ray electrons, increasing the likelihood of re-energized fossil plasma manifesting as radio phoenix emission.\\
Our multi frequency analysis reveals that certain emission regions, particularly the central region, are consistently detected across all frequencies, irrespective of the sensitivity. In contrast, features such as the \textbf{SW} extension and faint surrounding emissions are not visible at 1284 MHz, despite its superior sensitivity (see Fig. \ref{fig: mgcls and ugmrt}). This absence is likely due to significant synchrotron radiation and inverse Compton (IC) losses at higher frequencies. Additionally, the filament \textbf{NE2-N} undergoes distinct morphological changes across frequencies. It is completely absent at 887.5 MHz but observed at 1284 MHz as a broken tail. In particular, a blob-like structure is observed at the location of the second BCG (marked as F in Fig. \ref{fig: radio and optical} \textit{right}), as reported by \citetalias{slee2001four}, which also coincides with an X-ray peak (yellow '+' in Fig. \ref{fig: radio and optical} \textit{left}) identified by \citealt{juett2008chandra}. However, this feature cannot be conclusively identified as a radio galaxy, given its complete absence at 887.5 MHz and the lack of a corresponding point source in high-resolution radio images. Spectral curvature analysis reveals that this region exhibits near-zero curvature (see Fig. \ref{fig:curv map}), although the measurement carries significant uncertainty. Consequently, the nature of this emission remains ambiguous.\\
The central region was suggested as a past AGN activity of the NAT galaxy (Fig, \ref{fig: mgcls and ugmrt} \textit{right}, cyan contours within the magenta rectangular box of \textbf{peak1}). Taking account of its brightness distribution and morphology (eg., Source 1 in  Abell 1033, \citealt{de2015abell}), it resembles an FR-I galaxy with a host on the east side. It is also worth noting that FR-I galaxies are most probable to result as NAT and WAT, which are easily influenced by ICM. Moreover, the most likely host of this emission is the BCG. But from the Fig. \ref{fig: radio and optical} \textit{right}, it is found that the position of BCG is offset from the approximate head position of the NAT. \citealt{juett2008chandra} proposed two possible explanations based on the observation of a narrow filament, primarily outlining the NAT (see Sect. \ref{sec:A13}) and its westward extension. However, our observations reveal a more extended structure beyond this, indicating the need for an alternative origin scenario to account for the full extent of the diffuse emission.\\
One plausible alternative explanation is an adiabatic compression, wherein the radio cocoon less than 2 Gyr old following the cessation of AGN activity gets re-energized, as described in \citealt{ensslin2001reviving}. They have used three scenarios to define the re-energization mechanism for the formation of relic systems. Scenario A, where the radio cocoon remains at the cluster core, is relevant to our case. In this scenario, the  relatively high thermal pressure of the ICM (internal pressure of radio cocoon $P_{\mathrm{int}} \approx 2P_{\mathrm{ICM}}$) at the cluster center restricts adiabatic expansion, leading to un-decayed high internal magnetic field. As a result, the relativistic electrons experience rapid energy loss, shortening the duration of the Phase 2 (Lurking) stage. In this way, the cocoon becomes nearly undetectable and would require re-energization ideally within a spectral age of 0.1 Gyr. The proposed re-energization mechanism is via passage of shock. Due to higher internal sound speed, the shock is unable to penetrate, instead shock compresses the radio cocoon. This compression transforms the initially spherical cocoon into a toroidal structure, as demonstrated by simulations in \citealt{ensslin2002formation}. The simulation was done in four different cases, such as weak shock-weak magnetic field (\textit{wSwB}), weak shock-strong magnetic field (\textit{wSsB}) and so on. The weak shock is characterized by a compression factor of 2, and that of the strong shock is 3.3. Morphologically comparing, A13 radio phoenix emission is similar to the one as shown in Fig. 8 in the literature \citealt{ensslin2002formation}, which is an edge in view of a cocoon with a dynamically important strong magnetic field in compression via weak shock (\textit{wSsB}). This morphology exhibits two distinct regions of high surface brightness, likely caused by the compression of relativistic electrons and the resulting amplification of magnetic fields. They are similar to \textbf{Peak 1} and \textbf{Peak 2} of Fig. \ref{fig:A13_patch_label}. Despite the similarities, the filamentary structures (\textbf{NE1}, \textbf{NE2-S}, \textbf{NE2-N} in Fig. \ref{fig:A13_patch_label}) and the more extended faint emission observed at the periphery of this diffuse source remain unexplained under this adiabatic compression scenario. Moreover, the FR-I NAT jet structure extending from northwest to southeast remains obscure under this scenario  (Fig, \ref{fig: mgcls and ugmrt} \textit{right}. cyan contours  within the magenta rectangular box  of \textbf{peak1}). \\
An alternative explanation involves the role of ICM bulk motions. The shape of a relic system can trace not only the shocks but also the distribution of cosmic-ray (CR) electrons (\citealt{zuhone2021turning}). As a radio bubble or fossil AGN plasma rises buoyantly through a turbulent ICM, it can become distorted by Kelvin–Helmholtz and Rayleigh–Taylor instabilities (\citealt{zuhone2021turning}), resulting in the formation of disconnected pockets of coherent radio plasma. In this context, \textbf{Peak 2}, which lacks any known association with a radio galaxy and exhibits irregular morphology, could represent a second pocket related to \textbf{Peak 1}. The filamentary emissions observed in the \textbf{NE2-S} and \textbf{NE2-N} directions may also arise from redistributed relativistic electrons. As noted by \citealt{zuhone2021turning}, a re-energization mechanism is still necessary to make the fossil AGN plasma observable again. Similar re-energization has resulted in a radio phoenix in Abell 1033 (\citealt{de2015abell}). The cluster Abell 1033 was undergoing a merger in the north–south direction. A radio lobe, resembling FR-I NAT galaxy emission in the east–west direction, was found with its likely host located to the north, at a projected distance of 29 kpc. They have concluded that the radio lobe was revived by a shock detected toward the south via adiabatic compression. They proposed that the host had moved northward as the merger progressed. The emission showed a steep spectrum with a spectral index of $\alpha = -1.62$. The presence of a spectral bend was interpreted as the birth of a phoenix (\citealt{de2015abell}). 

The moderate mixing of relativistic electrons with different ages can be inferred from the patchy distribution of the spectral index (see Fig. \ref{fig:specmap}). The ultra-steep spectrum and negative spectral curvature depict the re-energization of aged relativistic electrons via adiabatic compression on shock passage (\citealt{2001A&A...366...26E}). Fossil AGN plasma loses energy primarily through synchrotron emission at mid-radio frequencies, while at higher frequencies, both synchrotron and inverse Compton (IC) losses dominate. These combined losses lead to a steepening and eventual curvature in the radio spectrum at high frequencies, resulting in the characteristic curved spectrum observed in fossil plasma. The compression shifts the radio spectrum to higher frequencies without altering the underlying electron energy distribution. Although no shocks were observed in the region of radio emission, this may be either due to the dense intracluster medium (ICM) in the cluster center, where shocks are typically weaker and become stronger toward the outskirts where radio gischt are observed (\citealt{van2019diffuse}), or due to projection effects, which can obscure shocks propagating at large angles with respect to the plane of the sky.
The high ICM pressure can also inhibit the free expansion of AGN radio lobes, causing them to remain confined (\citealt{parma2007search,murgia2011dying}). The undecayed higher internal magnetic field strength in such situations within the lobes results in increasing synchrotron emission and leads to the observed ultra-steep spectrum. This scenario is supported by the morphological signature of radio phoenix as its width widens toward regions of lower ICM density (see Fig. \ref{fig: radio and optical} \textit{left}). In the southwest extension, the expansion is likely more pronounced, which could weaken the magnetic field and lead to fainter emission, even at low frequencies. 
        
\section{Summary and Conclusion}\label{sec: summary}
Among the various classes of diffuse radio sources, radio phoenixes remain some of the least explored. However, existing studies have revealed several shared characteristics regarding their spatial distribution within clusters, origin, and spectral behavior. In this section, we summarize these common properties as discussed in \citet{mandal2020revealing} and relate them to our analysis of radio phoenix observed in Abell 13.
\begin{itemize}
    \item The radio phoenix, measured to have an LLS of 521 kpc at 400 MHz, begins at the cluster center and extends westward, where it reaches the outer boundary of the detected ICM X-ray emission region. (Fig. \ref{fig: radio and optical} \textit{left}). The $R_{500}$ of the cluster is 0.88 Mpc, which indicates that the emission is confined within the $R_{500}$.
    \item The source of seed relativistic electrons is likely supplied by past AGN activity, possibly associated with a NAT radio galaxy hosted by the BCG. 
    \item X-ray, optical, and radio observations report that the cluster is undergoing merging. The ICM bulk motion resulting from merger activity redistributes and mixes cosmic electrons with different ages. The patchy nature of the spectral index distribution shows that the mixing of relativistic electrons with different ages has occurred (Fig. \ref{fig: radio and optical}, \ref{fig:specmap}). The almost uniform distribution of negative spectral curvature indicate that the acceleration mechanism is common for all distinct regions of this emission (Fig. \ref{fig:curv map}).
    \item The integrated spectrum of the phoenix is well fitted by a second-order polynomial fit across the frequencies 147.5, 400, 887.5, and 1284 MHz. The fit yields an integrated spectral index of $-1.84 \pm 0.05$ at 400 MHz and a spectral curvature of $-0.93 \pm 0.21$ (Fig. \ref{fig:color-color and second order polynomial} \textit{right}). The observed spectral steepening and curvature at higher frequencies indicate that the AGN lobes, having previously lost energy through radiative cooling and inverse Compton (IC) scattering, have undergone re-energization, likely via adiabatic compression.
\end{itemize} 
Based on our analysis and the available multi-frequency radio data, we identify the diffuse emission in Abell 13 as a radio phoenix originating from fossil AGN radio plasma, likely associated with a NAT radio galaxy hosted by the BCG. The fossil plasma, originally injected during past AGN activity, appears to have been redistributed by bulk motion of the ICM  associated with cluster merger activity. This redistribution results in both an irregular morphology with filamentary structures and the mixing of relativistic electrons with different spectral ages. Over time, these electrons lose energy via adiabatic expansion and radiative processes, resulting in a steep radio spectrum. Subsequent re-energization, likely triggered by adiabatic compression by passage of a weak shock, revives the aged plasma, making the emission detectable again. The observed morphology, spectral behavior, and spatial association with the cluster center are consistent with the known characteristics of radio phoenixes. \\
To address the morphological and projection ambiguities of the radio phoenix, a polarization study is essential. Under the \textit{wSsB} edge-on view (Fig. 12; \citealt{ensslin2002formation}), compressed radio cocoons exhibit higher polarization fractions. Such measurements can reveal the magnetic field structure, offering insights into the formation mechanism and helping to resolve line-of-sight effects that obscure the true morphology and evolution of the emission.

\begin{acknowledgments}
We thank IIT Indore for providing the necessary computing facilities for data analysis. We acknowledge
the funding via the Department of Science and Technology, Government of India sponsored DST-FIST grant no. SR/FST/PSII/2021/162 (C).  Nasmi would like to thank DST for INSPIRE fellowship program for financial support (IF220302). SC acknowledges the support from Rhodes University and the National Research Foundation (NRF), South Africa. RR's research is supported by the South African Research Chairs Initiative of the Department of Science and Technology and the National Research Foundation (SARChI 81737).  MR acknowledges support from the National Science and Technology Council (NSTC) of Taiwan (NSTC 112-2628-M-007-003-MY3 and NSTC 114-2112-M-007-032-MY3) and the  Ministry of Education (MoE) Yushan Scholar program (MOE-108-YSFMS-0002-003-P1). This research has used the data available in the GMRT archives and archive images from MeerKAT (MGCLS) and ASKAP. We thank the staff of the GMRT who have made these observations possible. The GMRT is run by the National Center for Radio Astrophysics of the Tata Institute of Fundamental Research. The MeerKAT telescope is operated by the South African Radio Astronomy Observatory, which is a facility of the National Research Foundation, an agency of the Department of Science and Innovation. The Australian SKA Pathfinder
is part of the Australia Telescope National Facility which is managed by CSIRO.
The operation of ASKAP is funded by the Australian Government with support-
port from the National Collaborative Research Infrastructure Strategy. ASKAP
uses the resources of the Pawsey Supercomputing Centre. Establishment
of ASKAP, the Murchison Radio-astronomy Observatory and the Pawsey
Supercomputing Centres are initiatives of the Australian Government, with
support from the Government of Western Australia and the Science and
Industry Endowment Fund. We thank people from MeerKAT and ASKAP for providing us with the archive images. This research has used data
obtained from the \textit{Chandra} Data Archive. 
\end{acknowledgments}





%
\textit{Data Availability} : The archival radio data used in our work are available in the
GMRT Online archive (\hyperlink{https://naps.ncra.tifr.res.in/goa/data/
search}{https://naps.ncra.tifr.res.in/goa/data/search}).\\ The ASKAP archival image is available from \hyperlink{https://data.csiro.au/domain/casdaCutoutService}{https://data.csiro.au/domain/casdaCutoutService}. The archival MGCLS images are available from \url{https://archive-gw-1.kat.ac.za/public/repository/10.48479/7epd-w356/data/enhanced_products/index.html}

\facilities{GMRT, ASKAP, MeerKAT}

\software{CASA (\citealt{mcmullin2007casa}), 
          SPAM (\citealt{intema2009ionospheric,intema2014spam}), WSCLean (\citealt{offringa2014wsclean,offringa2017optimized}), CARTA \hyperlink{https://cartavis.org/}{https://cartavis.org/}, Astropy (\citealt{robitaille2013astropy,price2018astropy}), APLpy (\citealt{robitaille2012aplpy}), Matplotlib (\citealt{hunter2007matplotlib}), PyBDSF ( \citealt{mohan2015pybdsf})
          }

\clearpage
\appendix
\section{Spectral Index Low and Spectral Index High}\label{appendix:color}

\begin{table*}[!h]
    \begin{tabular}{|l|l|l|} 
    \hline
    \textbf{Cluster} & \textbf{Spectral Index Low} & \textbf{Spectral Index High} \\
     & \textbf{$\nu \in (150,325)$ MHz} & \textbf{$\nu \in (325,1400)$ MHz} \\
    \hline
    Abell 1914 & $-2.23$ & $-2.171$ \\
    SDSS-C4-DR3-3088 & $-1.81$ & $-2.09$ \\
    Abell 2593 & $-2.16$ & $-1.10$ \\
    Abell 2048 & $-1.35$ & $-2.3$ \\
    Abell 1033 & $-1.34$ & $-1.58$ \\
    Abell 2443 & $-2.83$ & $-2.84$ \\
    Abell 133 & $-1.24$ & $-2.05$ \\
    Abell 85$^{a}$ & $-1.83$ & $-2.48$ \\
    24P73 & $-1.58$ & $-2.41$ \\
    Ophiuchus & $-1.16$ & $-0.97$ \\
    Abell 2256 & $-0.66$ & $-1.94$ \\
    Abell 1931$^{b}$ & $-2.83$ & $-1.37$ \\
    MaxBCG J217.95869+13.53470 & $-1.09$ & $-2.80$ \\
    Abell 725 & $-0.85$ & $-0.86$ \\
    Abell 4038$^{c}$ & $-1.5$ & $-2.2$ \\
    Abell S753 & $-1.44$ & $-1.99$ \\
    CIZAJ1926.1+4833 & $-1.61$ & $-2.33$ \\
    MKW8 & $-1.84$ & $-3.23$ \\
    Abell 565 & $-1.28$ & $-2.05$ \\
    Abell 2675 & $-1.23$ & $-1.72$ \\
    Abell 272$^{d}$ & $-1.3$ & $-2.473$ \\
    Abell 566 & $-1.88$ & $-2.32$ \\
    Abell 2751 & $-0.77$ & $-1.34$ \\
    Abell 661 & $-0.62$ & $-1.06$ \\
    Abell 1550$^{e}$ & $-1.9$ & $-2.1$ \\
    Abell 2877$^{f}$ & $-2.16$ & $-4.87$ \\
    Abell 2256$^{g}$ & $-1.13$ & $-1.9$ \\
    Abell 655$^{h}$ & $-1.14$ & $-0.61$ \\
    \hline
    \end{tabular}

    \vspace{2mm}
    \footnotesize
    \parbox{\textwidth}{
    $^{a}$ $\alpha_{low} : \nu \in (148{-}323)$ MHz, $\alpha_{high} : \nu \in (323{-}1280)$ MHz \citealt{raja2023radio}.\\
    $^{b}$ $\alpha_{low} : \nu \in (143{-}325)$ MHz, $\alpha_{high} : \nu \in (325{-}1400)$ MHz \citealt{mandal2019ultra}.\\
    $^{c}$ $\alpha_{low} : \nu \in (74{-}327)$ MHz, $\alpha_{high} : \nu \in (327{-}1400)$ MHz \citealt{kale2018study}.\\
    $^{d}$ $\alpha_{low} : \nu \in (74{-}325)$ MHz, $\alpha_{high} : \nu \in (325{-}1400)$ MHz \citealt{whyley2024understanding}.\\
    $^{e}$ $\alpha_{low} : \nu \in (54{-}144)$ MHz, $\alpha_{high} : \nu \in (144{-}400)$ MHz \citealt{pasini2022particle}.\\
    $^{f}$ $\alpha_{low} : \nu \in (87{-}118)$ MHz, $\alpha_{high} : \nu \in (118{-}154)$ MHz \citealt{hodgson2021ultra}.\\
    $^{g}$ $\alpha_{low} : \nu \in (144{-}350)$ MHz, $\alpha_{high} : \nu \in (350{-}675)$ MHz \citealt{rajpurohit2022deep}.\\
    $^{h}$ $\alpha_{low} : \nu \in (20{-}40)$ MHz, $\alpha_{high} : \nu \in (40{-}144)$ MHz \citealt{groeneveld2025serendipitous}.
    }

    \caption{Spectral index values taken from the literature \citealt{mandal2020revealing}, except where noted.}
\end{table*}

\newpage
\bibliography{sample7}{}

\begin{thebibliography}{}
\expandafter\ifx\csname natexlab\endcsname\relax\def\natexlab#1{#1}\fi
\providecommand{\url}[1]{\href{#1}{#1}}
\providecommand{\dodoi}[1]{doi:~\href{http://doi.org/#1}{\nolinkurl{#1}}}
\providecommand{\doeprint}[1]{\href{http://ascl.net/#1}{\nolinkurl{http://ascl.net/#1}}}
\providecommand{\doarXiv}[1]{\href{https://arxiv.org/abs/#1}{\nolinkurl{https://arxiv.org/abs/#1}}}

\bibitem[{A. Botteon {et~al.}(2016)Botteon, Gastaldello, Brunetti, \& Kale}]{botteon2016shock}
Botteon, A., Gastaldello, F., Brunetti, G., \& Kale, R. 2016, \bibinfo{title}{A shock in ‘El Gordo’cluster and the origin of the radio relic,} Monthly Notices of the Royal Astronomical Society, 463, 1534

\bibitem[{H. Bourdin {et~al.}(2013)Bourdin, Mazzotta, Markevitch, Giacintucci, \& Brunetti}]{bourdin2013shock}
Bourdin, H., Mazzotta, P., Markevitch, M., Giacintucci, S., \& Brunetti, G. 2013, \bibinfo{title}{Shock heating of the merging galaxy cluster A521,} The Astrophysical Journal, 764, 82

\bibitem[{M. Brienza {et~al.}(2016)Brienza, Godfrey, Morganti, Vilchez, Maddox, Murgia, Orru, Shulevski, Best, Br{\"u}ggen, {et~al.}}]{brienza2016lofar}
Brienza, M., Godfrey, L., Morganti, R., {et~al.} 2016, \bibinfo{title}{LOFAR discovery of a 700-kpc remnant radio galaxy at low redshift,} Astronomy \& Astrophysics, 585, A29

\bibitem[{D.~S. Briggs(1995)Briggs}]{briggs1995high}
Briggs, D.~S. 1995, \bibinfo{title}{High fidelity deconvolution of moderately resolved sources,} Ph. D. Thesis

\bibitem[{M. Br{\"u}ggen {et~al.}(2018)Br{\"u}ggen, Rafferty, Bonafede, van Weeren, Shimwell, Intema, R{\"o}ttgering, Brunetti, Di~Gennaro, Savini, {et~al.}}]{bruggen2018discovery}
Br{\"u}ggen, M., Rafferty, D., Bonafede, A., {et~al.} 2018, \bibinfo{title}{Discovery of large-scale diffuse radio emission in low-mass galaxy cluster Abell 1931,} Monthly Notices of the Royal Astronomical Society, 477, 3461

\bibitem[{G. Brunetti \& T.~W. Jones(2014)Brunetti \& Jones}]{brunetti2014cosmic}
Brunetti, G., \& Jones, T.~W. 2014, \bibinfo{title}{Cosmic rays in galaxy clusters and their nonthermal emission,} International Journal of Modern Physics D, 23, 1430007

\bibitem[{G. Brunetti \& A. Lazarian(2016)Brunetti \& Lazarian}]{brunetti2016stochastic}
Brunetti, G., \& Lazarian, A. 2016, \bibinfo{title}{Stochastic reacceleration of relativistic electrons by turbulent reconnection: a mechanism for cluster-scale radio emission?} Monthly Notices of the Royal Astronomical Society, 458, 2584

\bibitem[{R. Cassano {et~al.}(2013)Cassano, Ettori, Brunetti, Giacintucci, Pratt, Venturi, Kale, Dolag, \& Markevitch}]{cassano2013revisiting}
Cassano, R., Ettori, S., Brunetti, G., {et~al.} 2013, \bibinfo{title}{Revisiting scaling relations for giant radio halos in galaxy clusters,} The Astrophysical Journal, 777, 141

\bibitem[{P. Chandra {et~al.}(2004)Chandra, Ray, \& Bhatnagar}]{chandra2004late}
Chandra, P., Ray, A., \& Bhatnagar, S. 2004, \bibinfo{title}{The late-time radio emission from SN 1993J at meter wavelengths,} The Astrophysical Journal, 612, 974

\bibitem[{S. Chatterjee \& A. Datta(2024)Chatterjee \& Datta}]{chatterjee2024deciphering}
Chatterjee, S., \& Datta, A. 2024, \bibinfo{title}{Deciphering the spectral properties of the atypical radio relic in A115 using uGMRT, VLA, and LOFAR,} Journal of Astrophysics and Astronomy, 46, 1

\bibitem[{S. Chatterjee {et~al.}(2022)Chatterjee, Rahaman, Datta, \& Raja}]{chatterjee2022unveiling}
Chatterjee, S., Rahaman, M., Datta, A., \& Raja, R. 2022, \bibinfo{title}{Unveiling the Origin of Peculiar Diffuse Radio Emission in A1351,} The Astronomical Journal, 164, 83

\bibitem[{C. De~Breuck {et~al.}(2002)De~Breuck, Tang, De~Bruyn, R{\"o}ttgering, \& van Breugel}]{de2002sample}
De~Breuck, C., Tang, Y., De~Bruyn, A., R{\"o}ttgering, H., \& van Breugel, W. 2002, \bibinfo{title}{A sample of ultra steep spectrum sources selected from the Westerbork In the Southern Hemisphere (WISH) survey,} Astronomy \& Astrophysics, 394, 59

\bibitem[{F. de~Gasperin {et~al.}(2015)de~Gasperin, Ogrean, van Weeren, Dawson, Br{\"u}ggen, Bonafede, \& Simionescu}]{de2015abell}
de~Gasperin, F., Ogrean, G., van Weeren, R., {et~al.} 2015, \bibinfo{title}{Abell 1033: birth of a radio phoenix,} Monthly Notices of the Royal Astronomical Society, 448, 2197

\bibitem[{F. De~Gasperin {et~al.}(2014)De~Gasperin, Van~Weeren, Br{\"u}ggen, Vazza, Bonafede, \& Intema}]{de2014new}
De~Gasperin, F., Van~Weeren, R., Br{\"u}ggen, M., {et~al.} 2014, \bibinfo{title}{A new double radio relic in PSZ1 G096. 89+ 24.17 and a radio relic mass--luminosity relation,} Monthly Notices of the Royal Astronomical Society, 444, 3130

\bibitem[{B. Dennison(1980)Dennison}]{dennison1980formation}
Dennison, B. 1980, \bibinfo{title}{Formation of radio halos in clusters of galaxies from cosmic-ray protons,} Astrophysical Journal, Part 2-Letters to the Editor, vol. 239, Aug. 1, 1980, p. L93-L96., 239, L93

\bibitem[{G. Di~Gennaro {et~al.}(2018)Di~Gennaro, Van~Weeren, Hoeft, Kang, Ryu, Rudnick, Forman, R{\"o}ttgering, Br{\"u}ggen, Dawson, {et~al.}}]{di2018deep}
Di~Gennaro, G., Van~Weeren, R., Hoeft, M., {et~al.} 2018, \bibinfo{title}{Deep Very Large Array observations of the merging cluster CIZA J2242. 8+ 5301: continuum and spectral imaging,} The Astrophysical Journal, 865, 24

\bibitem[{S. Duchesne {et~al.}(2021)Duchesne, Johnston-Hollitt, Offringa, Pratt, Zheng, \& Dehghan}]{duchesne2021diffuse}
Duchesne, S., Johnston-Hollitt, M., Offringa, A., {et~al.} 2021, \bibinfo{title}{Diffuse galaxy cluster emission at 168 MHz within the Murchison Widefield Array Epoch of Reionization 0-h field,} Publications of the Astronomical Society of Australia, 38, e010

\bibitem[{T.~A. Ensslin(2001)Ensslin}]{ensslin2001reviving}
Ensslin, T.~A. 2001, \bibinfo{title}{Reviving fossil radio plasma in clusters of galaxies by adiabatic compression in environmental shock waves,} Astronomy \& Astrophysics, 366, 26

\bibitem[{T.~A. Ensslin \& M. Brueggen(2002)Ensslin \& Brueggen}]{ensslin2002formation}
Ensslin, T.~A., \& Brueggen, M. 2002, \bibinfo{title}{On the formation of cluster radio relics,} Monthly Notices of the Royal Astronomical Society, 331, 1011

\bibitem[{T.~A. {En{\ss}lin} \&  {Gopal-Krishna}(2001){En{\ss}lin} \& {Gopal-Krishna}}]{2001A&A...366...26E}
{En{\ss}lin}, T.~A., \& {Gopal-Krishna}. 2001, \bibinfo{title}{{Reviving fossil radio plasma in clusters of galaxies by adiabatic compression in environmental shock waves},} \aap, 366, 26, \dodoi{10.1051/0004-6361:20000198}

\bibitem[{D. Fadda {et~al.}(1996)Fadda, Girardi, Giuricin, Mardirossian, \& Mezzetti}]{fadda1996observational}
Fadda, D., Girardi, M., Giuricin, G., Mardirossian, F., \& Mezzetti, M. 1996, \bibinfo{title}{The observational distribution of internal velocity dispersions in nearby galaxy clusters,} The Astrophysical Journal, 473, 670

\bibitem[{L. Feretti {et~al.}(1996)Feretti, B{\"o}hringer, Giovannini, \& Neumann}]{feretti1996radio}
Feretti, L., B{\"o}hringer, H., Giovannini, G., \& Neumann, D. 1996, \bibinfo{title}{The radio and X-ray properties of Abell 2255,} arXiv preprint astro-ph/9607027

\bibitem[{L. Feretti {et~al.}(2012)Feretti, Giovannini, Govoni, \& Murgia}]{feretti2012clusters}
Feretti, L., Giovannini, G., Govoni, F., \& Murgia, M. 2012, \bibinfo{title}{Clusters of galaxies: observational properties of the diffuse radio emission,} The Astronomy and Astrophysics Review, 20, 1

\bibitem[{Y. Fujita {et~al.}(2002)Fujita, Sarazin, Kempner, Rudnick, Slee, Roy, Andernach, \& Ehle}]{fujita2002chandra}
Fujita, Y., Sarazin, C.~L., Kempner, J.~C., {et~al.} 2002, \bibinfo{title}{Chandra observations of the disruption of the cool core in A133,} The Astrophysical Journal, 575, 764

\bibitem[{L.~T. George {et~al.}(2017)George, Dwarakanath, Johnston-Hollitt, Intema, Hurley-Walker, Bell, Callingham, For, Gaensler, Hancock, {et~al.}}]{george2017study}
George, L.~T., Dwarakanath, K., Johnston-Hollitt, M., {et~al.} 2017, \bibinfo{title}{A study of halo and relic radio emission in merging clusters using the Murchison Widefield Array,} Monthly Notices of the Royal Astronomical Society, 467, 936

\bibitem[{S. Giacintucci {et~al.}(2020)Giacintucci, Markevitch, Johnston-Hollitt, Wik, Wang, \& Clarke}]{giacintucci2020discovery}
Giacintucci, S., Markevitch, M., Johnston-Hollitt, M., {et~al.} 2020, \bibinfo{title}{Discovery of a giant radio fossil in the Ophiuchus galaxy cluster,} The Astrophysical Journal, 891, 1

\bibitem[{S. Giacintucci {et~al.}(2013)Giacintucci, Markevitch, Venturi, Clarke, Cassano, \& Mazzotta}]{giacintucci2013new}
Giacintucci, S., Markevitch, M., Venturi, T., {et~al.} 2013, \bibinfo{title}{New detections of radio minihalos in cool cores of galaxy clusters,} The Astrophysical Journal, 781, 9

\bibitem[{C. Groeneveld {et~al.}(2025)Groeneveld, van Weeren, Botteon, Cassano, de~Gasperin, Osinga, Brunetti, \& R{\"o}ttgering}]{groeneveld2025serendipitous}
Groeneveld, C., van Weeren, R., Botteon, A., {et~al.} 2025, \bibinfo{title}{Serendipitous decametre detection of ultra steep spectrum radio emission in Abell 655,} Astronomy \& Astrophysics, 693, A99

\bibitem[{T. Hodgson {et~al.}(2021)Hodgson, Bartalucci, Johnston-Hollitt, McKinley, Vazza, \& Wittor}]{hodgson2021ultra}
Hodgson, T., Bartalucci, I., Johnston-Hollitt, M., {et~al.} 2021, \bibinfo{title}{Ultra-steep-spectrum Radio “Jellyfish” Uncovered in A2877,} The Astrophysical Journal, 909, 198

\bibitem[{J.~D. Hunter(2007)Hunter}]{hunter2007matplotlib}
Hunter, J.~D. 2007, \bibinfo{title}{Matplotlib: A 2D graphics environment,} Computing in science \& engineering, 9, 90

\bibitem[{H. Intema(2014)Intema}]{intema2014spam}
Intema, H. 2014, \bibinfo{title}{SPAM: A data reduction recipe for high-resolution, low-frequency radio-interferometric observations,} arXiv preprint arXiv:1402.4889

\bibitem[{H. Intema {et~al.}(2009)Intema, Van~der Tol, Cotton, Cohen, Van~Bemmel, \& R{\"o}ttgering}]{intema2009ionospheric}
Intema, H., Van~der Tol, S., Cotton, W., {et~al.} 2009, \bibinfo{title}{Ionospheric calibration of low frequency radio interferometric observations using the peeling scheme-I. Method description and first results,} Astronomy \& Astrophysics, 501, 1185

\bibitem[{A.~M. Juett {et~al.}(2008)Juett, Sarazin, Clarke, Andernach, Ehle, Fujita, Kempner, Roy, Rudnick, \& Slee}]{juett2008chandra}
Juett, A.~M., Sarazin, C.~L., Clarke, T.~E., {et~al.} 2008, \bibinfo{title}{A Chandra observation of Abell 13: Investigating the origin of the radio relic,} The Astrophysical Journal, 672, 138

\bibitem[{R. Kale {et~al.}(2012)Kale, Dwarakanath, Bagchi, \& Paul}]{kale2012spectral}
Kale, R., Dwarakanath, K., Bagchi, J., \& Paul, S. 2012, \bibinfo{title}{Spectral and polarization study of the double relics in Abell 3376 using the Giant Metrewave Radio Telescope and the Very Large Array,} Monthly Notices of the Royal Astronomical Society, 426, 1204

\bibitem[{R. Kale {et~al.}(2018)Kale, Parekh, \& Dwarakanath}]{kale2018study}
Kale, R., Parekh, V., \& Dwarakanath, K. 2018, \bibinfo{title}{A study of spectral curvature in the radio relic in Abell 4038 using the uGMRT,} Monthly Notices of the Royal Astronomical Society, 480, 5352

\bibitem[{J.~C. Kempner {et~al.}(2003)Kempner, Blanton, Clarke, Ensslin, Johnston-Hollitt, \& Rudnick}]{kempner2003taxonomy}
Kempner, J.~C., Blanton, E.~L., Clarke, T.~E., {et~al.} 2003, \bibinfo{title}{A Taxonomy of Extended Radio Sources in Clusters of Galaxies,} arXiv preprint astro-ph/0310263

\bibitem[{S. Mandal(2020)Mandal}]{mandal2020revealing}
Mandal, S. 2020, PhD thesis, PhD thesis, Leiden Univ

\bibitem[{S. Mandal {et~al.}(2019)Mandal, Intema, Shimwell, Van~Weeren, Botteon, R{\"o}ttgering, Hoang, Brunetti, De~Gasperin, Giacintucci, {et~al.}}]{mandal2019ultra}
Mandal, S., Intema, H., Shimwell, T., {et~al.} 2019, \bibinfo{title}{Ultra-steep spectrum emission in the merging galaxy cluster Abell 1914,} Astronomy \& Astrophysics, 622, A22

\bibitem[{P. Mazzotta {et~al.}(2002)Mazzotta, Kaastra, Paerels, Ferrigno, Colafrancesco, Mewe, \& Forman}]{mazzotta2002evidence}
Mazzotta, P., Kaastra, J., Paerels, F., {et~al.} 2002, \bibinfo{title}{Evidence for a Heated Gas Bubble inside the “Cooling Flow” Region of MKW 3s,} The Astrophysical Journal, 567, L37

\bibitem[{J.~P. McMullin {et~al.}(2007)McMullin, Waters, Schiebel, Young, \& Golap}]{mcmullin2007casa}
McMullin, J.~P., Waters, B., Schiebel, D., Young, W., \& Golap, K. 2007, in Astronomical data analysis software and systems XVI, Vol. 376, 127

\bibitem[{N. Mohan \& D. Rafferty(2015)Mohan \& Rafferty}]{mohan2015pybdsf}
Mohan, N., \& Rafferty, D. 2015, \bibinfo{title}{Pybdsf: Python blob detection and source finder,} Astrophysics Source Code Library, ascl

\bibitem[{M. Murgia {et~al.}(2011)Murgia, Parma, Mack, De~Ruiter, Fanti, Govoni, Tarchi, Giacintucci, \& Markevitch}]{murgia2011dying}
Murgia, M., Parma, P., Mack, K.-H., {et~al.} 2011, \bibinfo{title}{Dying radio galaxies in clusters,} Astronomy \& Astrophysics, 526, A148

\bibitem[{A. Offringa \& O. Smirnov(2017)Offringa \& Smirnov}]{offringa2017optimized}
Offringa, A., \& Smirnov, O. 2017, \bibinfo{title}{An optimized algorithm for multiscale wideband deconvolution of radio astronomical images,} Monthly Notices of the Royal Astronomical Society, 471, 301

\bibitem[{A. Offringa {et~al.}(2014)Offringa, McKinley, Hurley-Walker, Briggs, Wayth, Kaplan, Bell, Feng, Neben, Hughes, {et~al.}}]{offringa2014wsclean}
Offringa, A., McKinley, B., Hurley-Walker, N., {et~al.} 2014, \bibinfo{title}{WSCLEAN: an implementation of a fast, generic wide-field imager for radio astronomy,} Monthly Notices of the Royal Astronomical Society, 444, 606

\bibitem[{A. Pal {et~al.}(2025)Pal, Kale, Wang, \& Wik}]{pal2025ugmrt}
Pal, A., Kale, R., Wang, Q.~H., \& Wik, D.~R. 2025, \bibinfo{title}{A uGMRT and MeerKAT Study of Radio Relics in the Low-mass Merging Cluster PSZ2 G200. 95- 28.16,} The Astrophysical Journal, 979, 4

\bibitem[{P. Parma {et~al.}(2007)Parma, Murgia, De~Ruiter, Fanti, Mack, \& Govoni}]{parma2007search}
Parma, P., Murgia, M., De~Ruiter, H., {et~al.} 2007, \bibinfo{title}{In search of dying radio sources in the local universe,} Astronomy \& Astrophysics, 470, 875

\bibitem[{T. Pasini {et~al.}(2022)Pasini, Edler, Br{\"u}ggen, De~Gasperin, Botteon, Rajpurohit, Van~Weeren, Gastaldello, Gaspari, Brunetti, {et~al.}}]{pasini2022particle}
Pasini, T., Edler, H., Br{\"u}ggen, M., {et~al.} 2022, \bibinfo{title}{Particle re-acceleration and diffuse radio sources in the galaxy cluster Abell 1550,} Astronomy \& Astrophysics, 663, A105

\bibitem[{G. Paturel {et~al.}(2003)Paturel, Petit, Prugniel, Theureau, Rousseau, Brouty, Dubois, \& Cambr{\'e}sy}]{paturel2003hyperleda}
Paturel, G., Petit, C., Prugniel, P., {et~al.} 2003, \bibinfo{title}{HYPERLEDA-I. Identification and designation of galaxies,} Astronomy \& Astrophysics, 412, 45

\bibitem[{S. Paul {et~al.}(2023)Paul, Kale, Datta, Basu, Sur, Parekh, Gupta, Chatterjee, Salunkhe, Iqbal, {et~al.}}]{paul2023exploring}
Paul, S., Kale, R., Datta, A., {et~al.} 2023, \bibinfo{title}{Exploring diffuse radio emission in galaxy clusters and groups with uGMRT and SKA,} Journal of Astrophysics and Astronomy, 44, 38

\bibitem[{R. Piffaretti {et~al.}(2011)Piffaretti, Arnaud, Pratt, Pointecouteau, \& Melin}]{piffaretti2011mcxc}
Piffaretti, R., Arnaud, M., Pratt, G., Pointecouteau, E., \& Melin, J.-B. 2011, \bibinfo{title}{The MCXC: a meta-catalogue of x-ray detected clusters of galaxies,} Astronomy \& Astrophysics, 534, A109

\bibitem[{A. Pinzke {et~al.}(2017)Pinzke, Oh, \& Pfrommer}]{pinzke2017turbulence}
Pinzke, A., Oh, S.~P., \& Pfrommer, C. 2017, \bibinfo{title}{Turbulence and particle acceleration in giant radio haloes: the origin of seed electrons,} Monthly Notices of the Royal Astronomical Society, 465, 4800

\bibitem[{A.~M. Price-Whelan {et~al.}(2018)Price-Whelan, Sip{\H{o}}cz, G{\"u}nther, Lim, Crawford, Conseil, Shupe, Craig, Dencheva, Ginsburg, {et~al.}}]{price2018astropy}
Price-Whelan, A.~M., Sip{\H{o}}cz, B., G{\"u}nther, H., {et~al.} 2018, \bibinfo{title}{The astropy project: Building an open-science project and status of the v2. 0 core package,} The Astronomical Journal, 156, 123

\bibitem[{R. Raja {et~al.}(2023)Raja, Rahaman, Datta, \& Smirnov}]{raja2023radio}
Raja, R., Rahaman, M., Datta, A., \& Smirnov, O.~M. 2023, \bibinfo{title}{A radio bridge connecting the minihalo and phoenix in the Abell 85 cluster,} Monthly Notices of the Royal Astronomical Society: Letters, 526, L70

\bibitem[{R. Raja {et~al.}(2020)Raja, Rahaman, Datta, Burns, Alden, Intema, van Weeren, Hallman, Rapetti, \& Paul}]{raja2020probing}
Raja, R., Rahaman, M., Datta, A., {et~al.} 2020, \bibinfo{title}{Probing the Origin of Diffuse Radio Emission in the Cool Core of the Phoenix Galaxy Cluster,} The Astrophysical Journal, 889, 128

\bibitem[{K. Rajpurohit {et~al.}(2022)Rajpurohit, Van~Weeren, Hoeft, Vazza, Brienza, Forman, Wittor, Dom{\'\i}nguez-Fern{\'a}ndez, Rajpurohit, Riseley, {et~al.}}]{rajpurohit2022deep}
Rajpurohit, K., Van~Weeren, R., Hoeft, M., {et~al.} 2022, \bibinfo{title}{Deep low-frequency radio observations of A2256. I. The filamentary radio relic,} The Astrophysical Journal, 927, 80

\bibitem[{S. Randall {et~al.}(2010)Randall, Clarke, Nulsen, Owers, Sarazin, Forman, \& Murray}]{randall2010radio}
Randall, S., Clarke, T., Nulsen, P., {et~al.} 2010, \bibinfo{title}{Radio and deep Chandra observations of the disturbed cool core cluster Abell 133,} The Astrophysical Journal, 722, 825

\bibitem[{C. Riseley {et~al.}(2024)Riseley, Bonafede, Bruno, Botteon, Rossetti, Biava, Bonnassieux, Loi, Vernstrom, \& Balboni}]{riseley2024meerkat}
Riseley, C., Bonafede, A., Bruno, L., {et~al.} 2024, \bibinfo{title}{A “MeerKAT-meets-LOFAR” study of the complex multi-component (mini-) halo in the extreme sloshing cluster Abell 2142,} Astronomy \& Astrophysics, 686, A44

\bibitem[{T. Robitaille \& E. Bressert(2012)Robitaille \& Bressert}]{robitaille2012aplpy}
Robitaille, T., \& Bressert, E. 2012, \bibinfo{title}{APLpy: astronomical plotting library in Python,} Astrophysics Source Code Library, ascl

\bibitem[{T.~P. Robitaille {et~al.}(2013)Robitaille, Tollerud, Greenfield, Droettboom, Bray, Aldcroft, Davis, Ginsburg, Price-Whelan, Kerzendorf, {et~al.}}]{robitaille2013astropy}
Robitaille, T.~P., Tollerud, E.~J., Greenfield, P., {et~al.} 2013, \bibinfo{title}{Astropy: A community Python package for astronomy,} Astronomy \& Astrophysics, 558, A33

\bibitem[{C.~L. Sarazin(2002)Sarazin}]{sarazin2002physics}
Sarazin, C.~L. 2002, \bibinfo{title}{The physics of cluster mergers,} Merging Processes in Galaxy Clusters, 1

\bibitem[{N. Seymour {et~al.}(2007)Seymour, Stern, De~Breuck, Vernet, Rettura, Dickinson, Dey, Eisenhardt, Fosbury, Lacy, {et~al.}}]{seymour2007massive}
Seymour, N., Stern, D., De~Breuck, C., {et~al.} 2007, \bibinfo{title}{The massive hosts of radio galaxies across cosmic time,} The Astrophysical Journal Supplement Series, 171, 353

\bibitem[{O. Slee \& J. Reynolds(1984)Slee \& Reynolds}]{slee1984steep}
Slee, O., \& Reynolds, J. 1984, \bibinfo{title}{Steep-Spectrum Radio Sources in Clusters of Galaxies—The Southern Sample,} Publications of the Astronomical Society of Australia, 5, 516

\bibitem[{O. Slee {et~al.}(2001)Slee, Roy, Murgia, Andernach, \& Ehle}]{slee2001four}
Slee, O., Roy, A., Murgia, M., Andernach, H., \& Ehle, M. 2001, \bibinfo{title}{Four extreme relic radio sources in clusters of galaxies,} The Astronomical Journal, 122, 1172

\bibitem[{R. Van~Weeren {et~al.}(2019)Van~Weeren, de~Gasperin, Akamatsu, Br{\"u}ggen, Feretti, Kang, Stroe, \& Zandanel}]{van2019diffuse}
Van~Weeren, R., de~Gasperin, F., Akamatsu, H., {et~al.} 2019, \bibinfo{title}{Diffuse radio emission from galaxy clusters,} Space Science Reviews, 215, 1

\bibitem[{T. Venturi {et~al.}(2022)Venturi, Giacintucci, Merluzzi, Bardelli, Busarello, Dallacasa, Sikhosana, Marvil, Smirnov, Bourdin, {et~al.}}]{venturi2022radio}
Venturi, T., Giacintucci, S., Merluzzi, P., {et~al.} 2022, \bibinfo{title}{Radio footprints of a minor merger in the Shapley Supercluster: From supercluster down to galactic scales,} Astronomy \& Astrophysics, 660, A81

\bibitem[{Z. Wang {et~al.}(2022)Wang, Murphy, Kaplan, Bannister, Lenc, Leung, O’Brien, Pintaldi, Pritchard, Stewart, {et~al.}}]{wang2022pilot}
Wang, Z., Murphy, T., Kaplan, D.~L., {et~al.} 2022, \bibinfo{title}{A pilot ASKAP survey for radio transients towards the Galactic Centre,} Monthly Notices of the Royal Astronomical Society, 516, 5972

\bibitem[{A. Whyley {et~al.}(2024)Whyley, Randall, Clarke, van Weeren, Rajpurohit, Forman, Edge, Blanton, Lovisari, \& Intema}]{whyley2024understanding}
Whyley, A., Randall, S.~W., Clarke, T.~E., {et~al.} 2024, \bibinfo{title}{Understanding the Nature of the Ultra-Steep Spectrum Diffuse Radio Source in the Galaxy Cluster Abell 272,} arXiv preprint arXiv:2402.04876

\bibitem[{J. ZuHone {et~al.}(2021)ZuHone, Ehlert, Weinberger, \& Pfrommer}]{zuhone2021turning}
ZuHone, J., Ehlert, K., Weinberger, R., \& Pfrommer, C. 2021, \bibinfo{title}{Turning AGN bubbles into radio relics with sloshing: modeling CR transport with realistic physics,} Galaxies, 9, 91

\end{thebibliography}
\bibliographystyle{aasjournalv7}



\end{document}